\documentclass[twocolumn]{aastex631}

\begin{document}

\title{Detection of GeV emission from an ultralong gamma-ray burst with the Fermi Large Area Telescope}
\author[0000-0002-6036-985X]{Yi-Yun Huang}
\affiliation{School of Astronomy and Space Science, Nanjing University, Nanjing 210093, China; hmzhang@nju.edu.cn; xywang@nju.edu.cn}
\affiliation{Key Laboratory of Modern Astronomy and Astrophysics (Nanjing University), Ministry of Education, Nanjing
210093, China}

\author[0000-0001-6863-5369]{Hai-Ming Zhang}
\affiliation{School of Astronomy and Space Science, Nanjing University, Nanjing 210093, China; hmzhang@nju.edu.cn; xywang@nju.edu.cn}
\affiliation{Key Laboratory of Modern Astronomy and Astrophysics (Nanjing University), Ministry of Education, Nanjing
210093, China}

\author[0000-0001-7542-5861]{Kai Yan}
\affiliation{School of Astronomy and Space Science, Nanjing University, Nanjing 210093, China; hmzhang@nju.edu.cn; xywang@nju.edu.cn}
\affiliation{Key Laboratory of Modern Astronomy and Astrophysics (Nanjing University), Ministry of Education, Nanjing
210093, China}
\author[0000-0003-1576-0961]{Ruo-Yu Liu}
\affiliation{School of Astronomy and Space Science, Nanjing University, Nanjing 210093, China; hmzhang@nju.edu.cn; xywang@nju.edu.cn}
\affiliation{Key Laboratory of Modern Astronomy and Astrophysics (Nanjing University), Ministry of Education, Nanjing
210093, China}
\author[0000-0002-5881-335X]{Xiang-Yu Wang}
\affiliation{School of Astronomy and Space Science, Nanjing University, Nanjing 210093, China; hmzhang@nju.edu.cn; xywang@nju.edu.cn}
\affiliation{Key Laboratory of Modern Astronomy and Astrophysics (Nanjing University), Ministry of Education, Nanjing
210093, China}

\begin{abstract}

GRB 220627A, detected by Fermi Gamma-ray Burst Monitor (GBM),  shows two episodes of gamma-ray emission, which are separated by a {$\sim$700} s long quiescent phase.
Due to similar temporal shapes and spectra in the two episodes, GRB 220627A is speculated to be a gravitationally lensed gamma-ray burst (GRB). 
We analyze the Fermi Large Area Telescope (LAT) data and find that about 49 gamma-ray photons above 100 MeV come from the GRB during the first episode, while there are no  photons above 100 MeV  in the second episode. Based on the broadband spectral study of the two episodes,  the gravitationally lensing scenario can be ruled out at a high confidence level and we thus conclude that GRB 220627A is an intrinsically ultralong GRB with the prompt burst emission lasting longer than 1000 s. 
It is then the first case that GeV emission is detected from an ultralong GRB. We find that a short spike seen in the LAT light curve  is also present in  GBM detectors that see the burst, suggesting a common internal region of emission across the entire Fermi energy range. The detection of a 15.7 GeV photon during the early prompt phase places a lower limit of $\Gamma\ge300$ on the bulk Lorentz factor of the GRB ejecta. The constraint on the bulk Lorentz factor could shed light on the origin of ultralong GRBs.

\end{abstract}


\section{Introduction}           
\label{sect:intro}

Gamma-ray bursts (GRBs) are the explosive phenomena in the universe and are usually divided into two classes based on their duration: long and short GRBs (e.g., \citealt{1981Ap&SS803M,Norris1984Natur.308..434N,Kouveliotou1993ApJ...413L.101K}). 
These objects have different progenitors, with the short (lasting less than 2\,s) GRBs believed to be formed in the merger of two compact objects (e.g., \citealt{Abbott2017ApJ848L12A} and long (lasting more than 2\,s) GRBs  originating from the core-collapse of massive stars (e.g., \citealt{Galama1998Natur.395..670G}). 
Recently, several works have proposed an additional category of ‘ultralong’ gamma-ray bursts (ULGRBs)\citep{Gendre2013ApJ,Stratta2013ApJ,Levan2014ApJ}, GRBs with duration of kiloseconds or longer.  
It has been proposed that the ULGRBs may have different progenitors than normal long-duration GRBs, such as blue supergiants \citep{Nakauchi2013ApJ,Levan2014ApJ,Perna2018ApJ}, white dwarf tidal disruption events (WD-TDEs; \citealt{Krolik2011ApJ,Levan2014ApJ,Ioka2016ApJ}),  and newborn magnetars \citep{Gompertz2017ApJ,Zou2019ApJ}.  
The blue supergiant model is a simple extension of the collapsar model for normal long GRBs, in which the ultralong duration arises from the much more extended envelope of the progenitor star, which leads to a much longer activity of the central engine \citep{Kashiyama2013ApJ,Nakauchi2013ApJ}. 
The WD-TDE model is  an extension of the model for jetted TDE Swift J1644+57 to slightly shorter duration \citep{Ioka2016ApJ}.  In the magnetar model, a rotating neutron star with a high surface magnetic field launches a magnetized jet for the duration of its spin-down time. 
Each progenitor model has distinct advantages and constraints when describing their observed properties. To date there is no smoking-gun observation of the progenitor of an ULGRB, and so each of these models remain plausible. 

At 21:21:00.09 UT on 2022 June 27, the Fermi Gamma-ray Burst Monitor (GBM) triggered and located GRB 220627A \citep{2022GCNgbm1,2022GCNgbm2}, which was also detected by the Large Area Telescope (LAT; \citealt{2022GCNlat}),  Konus-Wind \citep{2022GCNkw}, and Swift-BAT-GUANO \citep{2022GCNswift}. 
Separately, about 1000 s later at 21:36:56.39 UT, the GBM triggered, localized to a similar location, which was regarded as GRB 220627A triggered
once again by the GBM.  The duration ($T_{90}$) of the gamma-ray emission in the first episode is about 138 s (10--1000 \,keV), and that in the second episode is 127 s (10--1000 \,keV) \citep{2022GCNgbm2}. Due to the similar temporal shapes and spectra in the two episodes, GRB 220627A is speculated to be a  gravitationally lensed GRB \citep{2022GCNgbm2}. On the other hand, if the emission origin in the second episode is not due to gravitational lensing, GRB 220627A will be an ultralong GRB, given its extremely long duration.

Gravitational lensing produces multiple images of the same source, which differ in  intensity but have the same spectral shapes. 
Similarly, in time-varying sources, the temporal profile will be the same but
shifted in time for different images. 
For lensed GRBs,  lensed images are separated by milliarcseconds to arcseconds depending on the mass of the lens (e.g., millilensing events with mass $10^{4}-10^{7}M_{\odot}$ and marcrolensing events with mass $> 10^{7}M_{\odot}$ \citep{2010ARA&A..48...87T,lin2022ApJ...931....4L}, which cannot  be  resolved by current gamma-ray detectors. 
On the other hand, gamma-ray instruments have excellent time  resolution as well as sufficiently good spectral resolution \citep{Meegan2009ApJ}, and therefore the temporal and spectral information can be used to test the lensing scenario.  
The GRB lensing rates are highly uncertain, but somewhere on the order of one in a thousand GRBs \citep{mao1992ApJ}. 
Recently, there has been an increase in claims of lensed GRB events. \cite{Paynter2021NatAs...5..560P} presented evidence for lensing in the short-duration BATSE GRB 950830. 
\cite{Wang2021ApJ...918L..34W} and \cite{Yang2021ApJ...921L..29Y} independently proposed that the short GRB, 200716C, shows lensing signatures. 
Based on the analysis of the autocorrelation function, \cite{Kalantari2021ApJ...922...77K} claimed a  lensing candidate, the long-duration GRB 090717. \cite{Veres2021ApJ...921L..30V} found that  the two emission episodes of GRB 210812A have the same pulse and spectral shape,  favoring the lensing origin. \cite{lin2022ApJ...931....4L} present a systemic search for millilensing among 3000 GRBs observed by Fermi-GBM up to 2021 April and find four interesting candidates.

In this Letter, we first study whether the two separate emission episodes could be due to gravitational lensing for GRB 220627A. As this GRB is observed by both GBM and LAT, we use the broadband spectral analysis to test the lensing scenario. We rule out the lensing scenario based on the large difference in the flux ratio between the GeV and MeV emission for the two episodes. Thus, GRB 220627A is a truly ultralong GRB with a duration longer than 1000 s. Then our detection at GeV band of GRB~220627A represents the first case that GeV photons are detected from ULGRBs. As the detection of GeV emission implies a high bulk Lorentz factor for the GRB jet, this could be useful to constrain the origin of ULGRBs.

Throughout this Letter, we adopt a Hubble constant of $H_{0} = 71 ~\rm {km~s^{-1}~ Mpc^{-1}}$ and cosmological parameters of  $\Omega_{\rm m}=0.27$, and $\Omega_{\Lambda}=0.73$.

\section{Data Reduction and Analysis}
\subsection{GBM Data Reduction}
Fermi-GBM triggered and located GRB 220627A at 21:21:00.09 UT ($T_0$) on 2022 Jun 27 \citep{2022GCNgbm1,2022GCNgbm2}. It was also detected by Fermi-LAT, which was 27$\degr$from the LAT boresight at the time of the GBM trigger \citep{2022GCNlat}. Note that, about 1000 s later at 21:36:56.39 UT, GBM triggered and located to a similar location, which was regarded as GRB 220627A triggered once again by the GBM \citep{2022GCNgbm2}.
GBM has 12 sodium iodide (NaI) and two bismuth germanate (BGO) scintillation detectors, covering the energy range 8 keV$-$40 MeV \citep{Meegan2009ApJ}. 
We downloaded GBM data of this GRB from the Ferm-GBM public data archive.\footnote{\url{https://heasarc.gsfc.nasa.gov/FTP/fermi/data/gbm/daily/}}
For the first trigger, the detectors selected for our analysis are: two NaI detectors (namely $n0$ and $n3$) and one BGO detector (namely b0), which have the smallest viewing angles with respect to the GRB source direction.
For the second trigger, two NaI detectors, $n6$ and $n7$ and one BGO detector, b1, are selected for our analysis.

\subsection{LAT Data Reduction}

The Fermi-LAT extended type data for the GRB 220627A were taken from the Fermi Science Support Center\footnote{\url{https://fermi.gsfc.nasa.gov}}. 
Only the data within a $14\degr \times14\degr$ region of interest (ROI) centered on the position of GRB 220627A are considered for the analysis (initially centered on the Ferm-GBM  position).

We perform an unbinned maximum likelihood analysis, using LAT $TRANSIENT$ events between 100 MeV and 30 GeV, and with a maximum zenith angle of 100$\degr$ to reduce the contamination from the $\gamma$-ray Earth limb.
We select a time interval of 0--700 s after the GBM trigger time T0, which contains all the gamma-rays detected by LAT before the GRB exited its field of view (FOV). 
The instrument response function (IRF) (\textit{$P8R3\_TRANSIENT020\_V$}3)\footnote{\url{https://fermi.gsfc.nasa.gov/ssc/data/analysis/documentation/Cicerone/Cicerone_Data/LAT_DP.html}} is used.
The main background component consists of charged particles that are misclassified as gamma-rays. It is included in the analysis using the isotropic emission template (``$iso\_P8R3\_TRANSIENT020\_V3\_v1.txt$'').
The contribution from the Galactic diffuse emissions is accounted for by using the diffuse Galactic interstellar emission template (IEM; $gll\_iem\_v07.fits$). The parameter of isotropic emission and IEM are left free.

Assuming a power-law spectrum ($dN/dE = AE^{\Gamma}$) of the burst, we obtained the best-fit Fermi-LAT position of GRB 220627A in 0--700 s with the tool \emph{gtfindsrc}: (201.$\degr$22, -32.$\degr$49) with a circular error of 0.$\degr$02 (statistical only), which is consistent with the result reported by \cite{2022GCNlat}.

The maximum likelihood test statistic (TS) is used to estimate the significance of the GRB, which is defined by TS$= 2 (\ln\mathcal{L}_{1}-\ln\mathcal{L}_{0})$, where $\mathcal{L}_{1}$ and $\mathcal{L}_{0}$ are maximum likelihood values for the background with the GRB and without the GRB (null hypothesis).
The TS value for GRB 220627A is found to be 182.89 (corresponds to a detection
significance of $13.52\sigma$) in 0--700 s, and the averaged flux is $(1.17\pm0.29)\times 10^{-8}\ {\rm erg \ cm^{-2} \ s^{-1}}$ with a photon index $\Gamma_{\rm LAT}=-2.16\pm0.15$. Assuming the redshift of the GRB is $z=3.084$ \citep{2022GCN.32291....1I}, the measured value of the flux corresponds to a luminosity of $(1.01\pm0.25)\times 10^{51}\ {\rm erg \ s^{-1}}$.

As part of our analysis, we use the \emph{gtsrcprob} tool to estimate the probability that each photon detected by the LAT is associated with the GRB.
There are 86 events detected by LAT before the GRB exited its FOV, and the 47 events are associated with the GRB with a probability greater than 80\%, among which 35 photons have a probability higher than 90\%. 
The highest-energy photon is a 15.73 GeV photon arriving at $ T_{0}+176\ s$, which has a probability $\sim99.9\% $ associated with this GRB.

\section{Light Curves}

\begin{figure*}
\includegraphics[angle=0,scale=0.45]{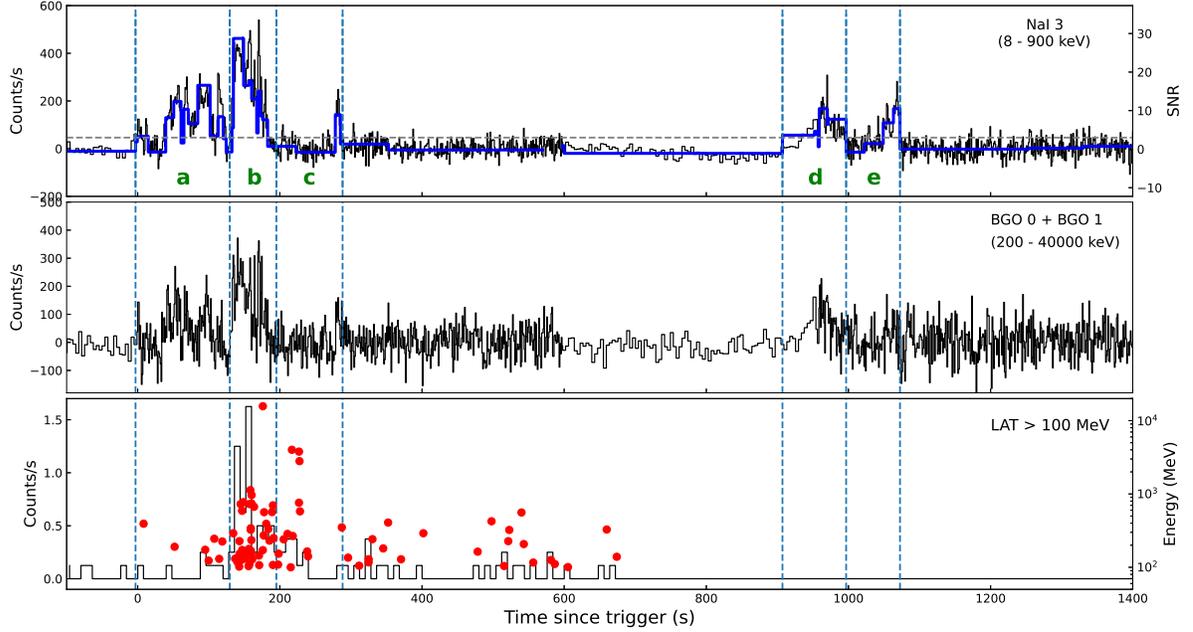}
\caption{GBM and LAT light curves of GRB 220627A.
The background subtracted light curves of NaI and BGO are extracted from the CSPEC type data, which use a 1024 ms time bin after the detector trigger and a 4096 ms time bin before the detector trigger. The blue solid line represents the Bayesian block light curve and the gray dashed line represents that the S/R is equal to 3.
The last panel shows the LAT TRANSIENT class events for energies $>$100 MeV using 10 s time bins. The red points show the arrival time and corresponding energy of LAT photons.
The vertical dashed lines indicate the time intervals for the time-resolved spectral analysis derived from the Bayesian block: $T_0$ +(-3.08, 129.53, 195.07, 288.26, 907.15, 996.75, 1072.53) s.  }
\label{multiLC}
\end{figure*}

In Figure \ref{multiLC}, we show the GBM and LAT light curves in several energy bands.
The GBM light curve, derived from CSPEC\footnote{\url{https://fermi.gsfc.nasa.gov/ssc/data/access/gbm/}} type data in the energy range 8--900 keV, shows multiple bursts over 1000 s, and it can be separated into two emission episodes. 
For the first episode, the light curves show multiple peaks with a duration of $T_{90} = 134.09 \rm s$ in 8--900 keV. The emission measured by  LAT has a  spike that is also seen in  GBM detectors that see the burst, suggesting a common internal region of emission across the entire Fermi energy range. The presence of a similar sharp peak
in both GBM and LAT has  been seen in GRB 090926A \citep{Ackermann2011ApJ}.
At $T_0 +956.3$ s, the GBM triggered once again and the light curves of this episode show a similar shape to that of the first episode, with a duration of $T_{90}= 118.66 \rm \ s$ in 8--900 keV. But LAT did not detect any high-energy photons in this episode, even though the burst was still within the field of view of LAT with a boresight angle of 56.$\degr$8.
To determine the time intervals  for spectroscopy analysis, we employ the Bayesian block  method \citep{2013ApJ...764..167S} on the photon events between 8 and 900 keV, which are  indicated by the vertical lines in Figure \ref{multiLC}, with boundaries shown in Table \ref{tab:intervals}.
The background is modeled via applying the “baseline” method \citep{2018pgrb.book.....Z} to a wide time interval around the signal and has been subtracted in GBM light curves. 
The resulting blocks, as shown in the upper panel of Figure \ref{multiLC}, can track the statistically significant changes in the light curve. The minimum bin size of the obtained blocks is 1024 ms. 

In order to quantify the fluctuations of temporal profiles, we also employ the Bayesian block  method on the unbinned event data of 8$-$900 keV. We take half of the minimum block size as the minimal variability time for this burst (e.g., \citet{Vianello2018ApJ}), which yields a minimal variability time of $\delta t\simeq 1.79 \rm \,s$ for the first episode. 
 
\begin{figure*}
\includegraphics[angle=0,scale=0.8]{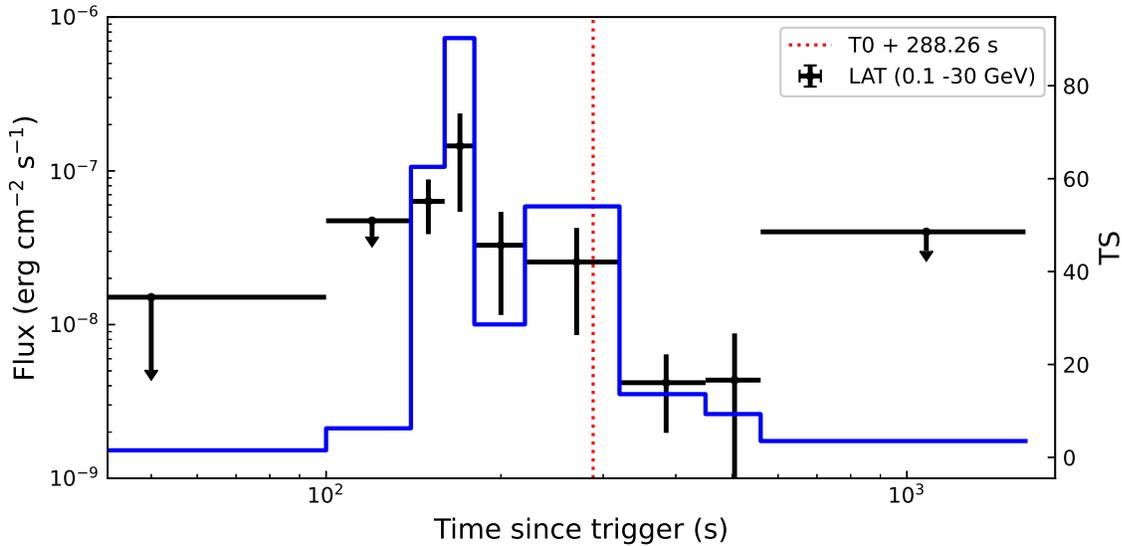}
\caption{Long-term LAT light curve of GRB 220627A. The black data points represent the GeV light curve of GRB 220627A, while the arrows show the upper limits when the TS value of the data points is less than nine. The blue line shows the TS value of the corresponding data points. The vertical dashed line represents the end of the first emission episode measured by GBM.}
\label{LAT_LC}
\end{figure*}

For the long-time analysis of LAT emission, we performed an unbinned likelihood analysis of the LAT $>100$ MeV data, and the light curve is shown in Figure \ref{LAT_LC}. Apparently, after the prompt emission phase (up to $T_0+$ 288.26 s), weak GeV emission is present but no obvious keV$-$MeV emission are found in the Figure \ref{multiLC}, which may be simply interpreted as arising from the afterglow emission \citep{2009MNRAS.400L..75K,2010ApJ...712.1232W,2011ApJ...733...22H} .

\section{Spectral analysis}

We first performed a time-integrated joint spectral analysis of the LAT and GBM data for the two emission episodes. 
The data for Fermi-GBM spectra were extracted using the public software \textit{gtburst}\footnote{\url{https://fermi.gsfc.nasa.gov/ssc/data/analysis/scitools/gtburst.html}}. We analyzed the data of the two NaI detectors with the smallest viewing angle ($n0$, $n3$ for the first episode and n6, n7 for the second) and the most illuminated BGO detector (b0 or the first episode and b1 for the second). We selected the energy channels in the range 8--900 keV for NaI detectors, excluding the channels in the range 32--36 keV (due to the iodine K-edge at 33.17 keV), and 0.2--40 MeV for the BGO detector. 
For the background modeling, we manually selected two time intervals (one before the burst and the other after the burst), fitting these with a polynomial function whose order is automatically found by \textit{gtburst}.
We first use the Band function and cutoff power-law (CPL) function  to fit the time-integrated spectra. 
The Band model can be expressed as
\begin{equation}
 N(E)=\left\{
 \begin{array}{l}
 A(\frac{E}{100\,{\rm keV}})^{\alpha}{\rm exp}(-\frac{E}{E_{\rm c}}),\,E<(\alpha-\beta)E_{\rm c} \\
 A\big[\frac{(\alpha-\beta)E_{\rm c}}{100\,{\rm keV}}\big]^{\alpha-\beta}{\rm exp}(\beta-\alpha)(\frac{E}{100\,{\rm keV}})^{\beta}, E\geq(\alpha-\beta)E_{\rm c} \\
 \end{array}\right.
\end{equation}
where $\alpha$ and $\beta$ are low-energy and high-energy photon spectral indices. The peak energy $E_{\rm p}$ is related to the cutoff energy, $E_{\rm c}$, through $E_{\rm p}=(2+\alpha)E_{\rm c}$. 
The CPL model is expressed as 
\begin{equation}
 N(E)=AE^{-\lambda}{\rm exp}(-E/E_{\rm c}),
\end{equation}
where $\lambda$ is the power-law photon index below the cutoff energy and $E_{\rm c}$ is $e$-folding energy. The joint spectral fitting of GBM and LAT data was performed using XSPEC v12.12.0 \citep{xspec1996ASPC}, and C-stat is used as the statistic to estimate uncertainties of the best-fit parameters. 

\begin{figure*}
\centering{}
\includegraphics[angle=0,scale=0.56]{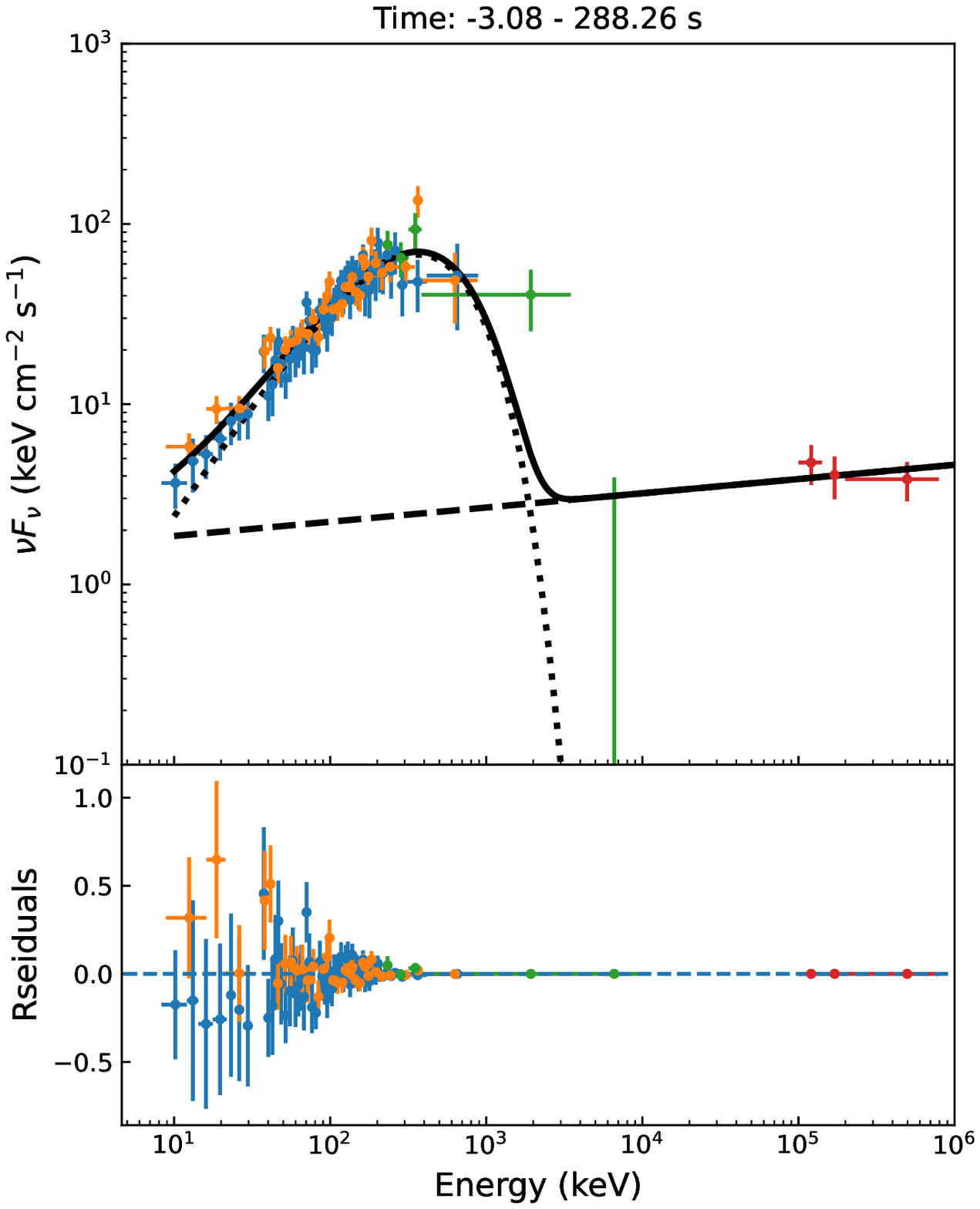}
\includegraphics[angle=0,scale=0.56]{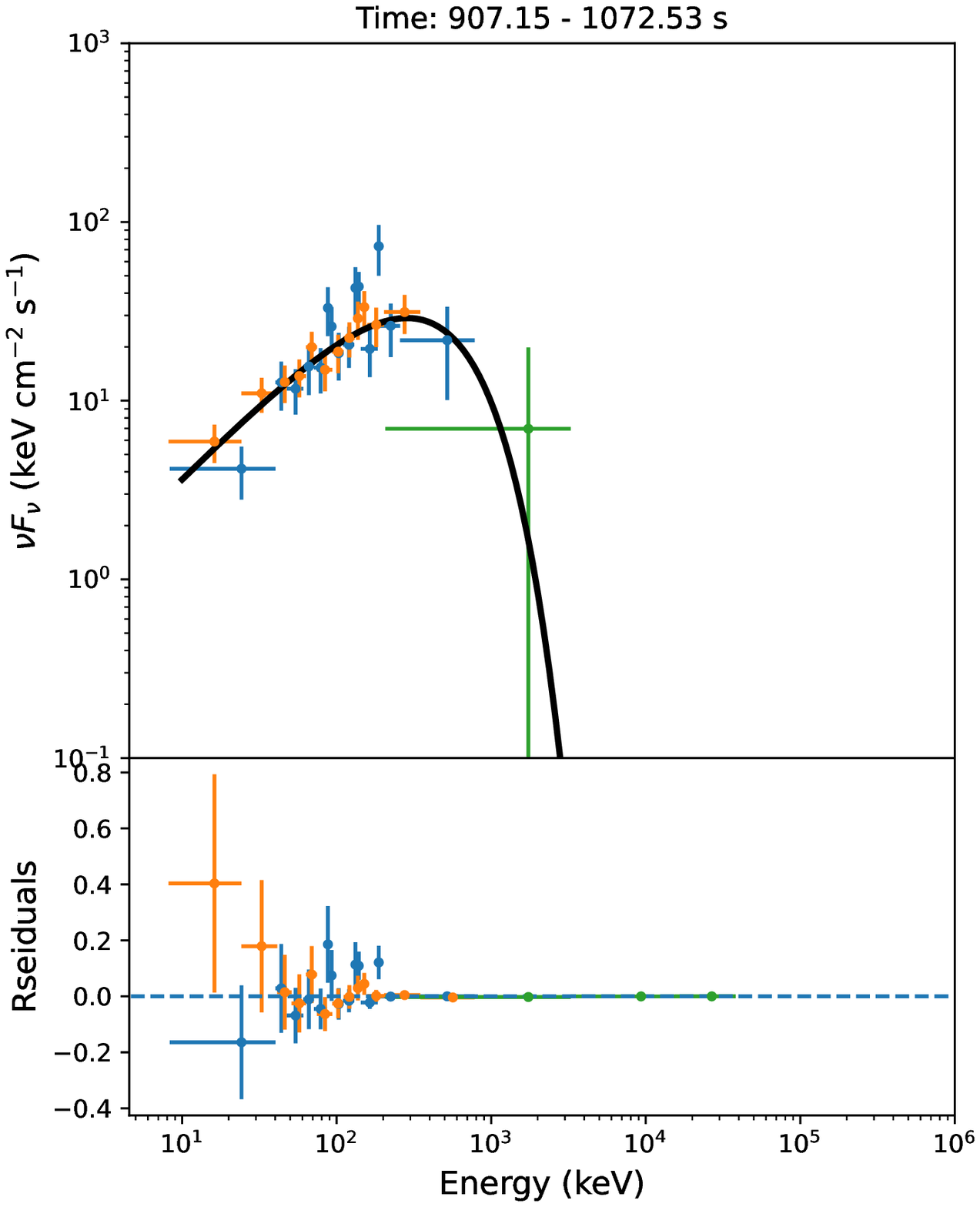}
\caption{Broadband spectrum of  of GRB 220627A in the two emission episodes. 
Left panel: the time-integrated spectrum measured from $T_{0}-3.08 \rm \,s$ to $T_{0}+288.53 \rm \,s$ and the fitting with the CPL+PL model.  The dotted and dashed lines represent the CPL component and the PL component, respectively, and the solid  lines represent the sum of them.
Right panel: the time-integrated spectrum measured from $T_{0}+907.15 \rm \,s$ to $T_{0}+1072.53 \rm \,s$ and the fitting with the CPL model.
}
\label{sed0}
\end{figure*}

Table \ref{tab:1} summarizes the best-fit parameters for the testing models. As seen in Table \ref{tab:1}, the CPL model cannot describe the data well and we then find that adding an extra power-law component can significantly improve the fit statistics.
We employ Bayesian information criterion (BIC; \citealt{1978AnSta...6..461S}) to compare models, which is a method taking into account the model complexity and the different numbers of free parameters. We find that CPL+PL model provide a statistical improvement over the simpler Band function and CPL model during the first episode, with $\Delta_{\rm BIC}=  7.89 $ and $\Delta_{\rm BIC}= 1479.46$, respectively\footnote{BIC is defined as $BIC = \chi^2 + klnN$, where $\chi^2$ is the fit statistic, N is the number of data points, and k is the number of free parameters of the model. { The strength of the evidence
against the model with the higher BIC value can be summarized as follows \citep{2017JCAP...01..005N}. (1)
If $\Delta_{\rm BIC}\geq 2$, there is no evidence against the higher BIC model; (2) if $4\leq \Delta_{\rm BIC}\leq 7$, positive evidence against the higher BIC model is given;
(3) if $\Delta_{\rm BIC} \geq 10$, very strong evidence against the higher BIC model is given.}}, indicating no support for the Band function or CPL model \citep{2017JCAP...01..005N}. 
Also, we test the Band+PL model and find no statistical improvement over the simple Band function. Therefore, the CPL+PL model is the best-fit model for the first episode.
The parameters of the CPL+PL model are $\lambda = 0.73\pm0.03$ and the cutoff energy is $E_c = 286.96\pm38.00 \, \rm keV$ for the CPL component, and it has an index of $1.92\pm0.05$ for the PL component. 
The fit results are summarized in Table \ref{tab:1}, and the spectra and residuals are shown in Figure \ref{sed0} for the best-fit model.


For the second episode, we only consider the Band function and CPL model since there are no GeV photons detected by LAT during this episode. We find that the second episode is described better by a CPL model than a Band function ($\Delta_{\rm BIC}$ = 5.84), since the parameter $\beta$ of the Band function cannot be well constrained. The best-fit parameters of the CPL model are $\lambda = 1.06\pm0.11$ and  $E_c = 248.64\pm66.58 \, \rm keV$. 
The fitting results are summarized in Table \ref{tab:1}, and the spectra and residuals are shown in the right panel of Figure \ref{sed0} for the best-fit model.

Then, we performed a time-resolved joint spectral analysis for the time interval $a$, $b$, $c$, $d$ and $e$ to study the potential spectral evolution. As shown in Figure \ref{multiLC}, interval (a) represents the beginning of the prompt phase, which can be better described by the Band function like other GRBs with
the parameters: $\alpha=-0.89 \pm 0.04$, $\beta=-2.45 \pm 0.02$, and $E_{p}=334.08 \pm 33.32 \rm \ keV$. 
During interval (b), which is the most strong emission phase in the first episode, the emission spectrum is best described by a cutoff PL plus an extra power-law component, similar to the time-integrated spectrum with the parameters: $\lambda = 0.64\pm0.10$, $E_c = 267.35\pm34.75 \, \rm keV$, and a hard power-law index $1.84\pm0.06$.
As shown in the upper panel of Figure \ref{multiLC}, interval (c) has only one Bayesian block with signal to noise (SNR) greater than 3. For the sake of a complete analysis of the first episode,   we also do the fitting of this interval. We find that the parameters of tested model for this interval can not be constrained well.
Parameters of the preferred model for interval c are: $\alpha=-1.79\pm0.12$, $\beta=-2.26\pm0.46$, and $E_{p}=9748.51\pm6635.92 \rm \ keV$. 
For the time intervals (d) and (e), the CPL model is the preferred model, which are similar to the time-integrated spectrum. The fitting results is summarized in Table \ref{tab:intervals}.

 The GeV emission detected by LAT during the first episode belongs to the prompt emission, as suggested by  a common  spike seen in both the LAT  and GBM  detectors. The  power-law spectrum of this GeV emission indicates that it is not an extrapolation of the Band component, but is instead  an extra power-law component, which has been seen in some other GRBs, such as GRB 090926A \citep{Ackermann2011ApJ}. This  extra power-law component disappears in the second episode, indicating that the component evolves quickly, which is consistent with the hard-to-soft evolution seen during the prompt emission of GRBs.

\section{Implication for the lensed GRB scenario}

The broadband spectrum of the first episode emission is well described by a CPL plus an extra hard PL component. The necessity of the extra power-law component is due to the presence of the hard GeV emission spectrum. 
The averaged photon flux of GRB 220627A in the LAT energy band is  $(2.88\pm0.47)\times 10^{-5}\ {\rm photons \ cm^{-2} \ s^{-1}}$ during the first episode. By considering the corresponding exposure time and effective area of LAT in this episode, there are $48.53\pm7.89$  photons above 100 MeV that are detected from GRB 220627A.
On the other hand, during the second episode,  no photons above 100 MeV were detected by LAT. The emission spectrum during the second episode is well described by a CPL model. 

Assuming that the second  episode emission has the same spectral form as that of the first episode, as motivated by the lensing scenario, we expect that $5.98\pm0.97$ photons above 100 MeV should have been detected by LAT. 
We estimate the chance probability of detecting gamma-ray photons with $N<1$ ($N$ is the total number of photon from the GRB detected by LAT) to be $1.42\times10^{-7}$, suggesting that the lensing scenario can be ruled out at a confidence level of $5.1\sigma$.

We can also test the lensing scenario by comparing interval (b) and (d), since the emission in these two intervals would result from the lensing of a single pulse. During interval (b), $33.43\pm6.32$ gamma-ray photons above 100 MeV were detected from GRB 220627A by LAT. Assuming that the emission in interval (d) has the same spectral form as that of interval (b),  we expect that $3.83\pm0.72$ photons above 100 MeV should have been detected by LAT during interval (d). The fact that no gamma-ray photons are detected also strongly disfavors the lensing scenario.
The chance probability of detecting less than one gamma-ray photon from the GRB in this episode is $4.24\times10^{-5}$,  suggesting that the lensing scenario is excluded at a  confidence level of $3.9\sigma$.

\section{Constraints on the bulk Lorentz factor}

The bulk Lorentz factor ($\Gamma$) of the ejecta is an essential  parameter to understanding the physics of GRBs. The prompt emission of GRBs is thought to be produced in an
ultrarelativistic ejecta, as argued by the fact that  high-energy photons ($\gg {\rm MeV}$) escape out of the source without
suffering from absorption due to pair production
($\gamma\gamma\rightarrow e^+e^-$) (e.g., \citealt{Krolik1991ApJ,Fenimore1993A&AS,2001ApJ...555..540L,2015ApJ...806..194T}).
Requiring that the absorption optical depth
$\tau_{\gamma\gamma}\la1$ within the source for high energy photons, one can obtain
a lower limit on  the bulk Lorentz factor ($\Gamma$) of the
emitting region.

We consider a simple one-zone model for GRB 220627A where the high-energy photons come from the same region as the low-energy target photons, as suggested by  a common  spike seen in both the LAT  and GBM  detectors, as shown in Figure \ref{multiLC}. The same structure is also seen in some other GRBs, such as GRB 090926A \citep{Ackermann2011ApJ}.
The target photons that annihilate with photons of energy $E_M$ should have energy above $E_t =2\Gamma^2(m_e c^2)^2/[E_M(1+z)^2]$, where $E_M$ is the maximum energy of the photons detected by LAT and $z$ is the redshift of GRB 220627A. These photons usually
come from the high-energy part of the prompt emission spectrum. Since the Band function  describes the data of GRB 220627A almost equally well as the cutoff PL function plus an extra power-law component, we replace the spectral form with the Band function in computing the absorption depth  for the convenience of calculation.  
The $\gamma\gamma$ optical depth \textbf{($\tau_{\gamma\gamma}$)} can be given by \cite{2008ApJ...677...92G,2012MNRAS.421..525H,2018MNRAS.478..749C}
\begin{equation}
\begin{array}{ll}
\tau_{\gamma\gamma}(E_M)=2^{1+2\beta}I(\beta)\frac{(-2-\beta)\sigma_{\rm T}L_{\gamma}(E>E_{\rm peak})\delta t\Gamma^{2(1+\beta)}}{4\pi R^2E_{\rm peak}(1+z)} \\ \times
\left[\frac{m_e^2c^4}{E_M E_{\rm peak}(1+z)^2} \right]^{1+\beta}
\end{array}
\end{equation}
where  $\sigma_{\rm T}$ is the Thomson cross section, and $\beta$ is the high-energy spectral index. The radius of the fireball is given by $R=2\Gamma^2c\delta t$ with $\delta t\,$($\simeq 1.79$ s) being the variability timescale of a single pulse; $I(\beta)=\int_0^1yg(y)dy/(1-y^2)^{2+\beta}$ with $g(y)=\frac{3}{16}(1-y^2)\left[(3-y^4){\rm ln}\frac{1+y}{1-y}-2y(2-y^2)\right]$. Here we assume that only photons with energy beyond the spectral $E_{\rm peak}$ are energetic enough to annihilate with high-energy photons, and hence the low-energy spectral slope $\alpha$ does not show up in the expression. 
The maximum energy of the LAT detected photons in GRB 220627A is $E_M=15.73$ GeV. From the above equation, we obtain a lower limit on the bulk Lorentz factor of $\Gamma\ge 300$ for GRB 220627A.

\section{Discussions and Conclusions}
Various  models have been proposed for the origin of ULGRBs, including collapsing blue supergiants \citep{Nakauchi2013ApJ,Levan2014ApJ,Perna2018ApJ}, which  have larger envelopes than normally considered for GRB progenitors, tidal disruption of a white-dwarf star by a massive
black hole \citep{Krolik2011ApJ,Levan2014ApJ,Ioka2016ApJ}, and newborn millisecond magnetars  \citep{Gompertz2017ApJ,Zou2019ApJ}. It is unclear how large the bulk Lorentz factors of jets can be produced in the above models. But the bulk Lorentz factors could, in principle,  be largely different in these models, as the baryon loading in these models may be quite different. The WD-TDE model was originally proposed to explain the jetted TDE events, such as 
Swift J1644+57. The null result of the search for  GeV photons from these jetted  TDEs could be due to the fact that the high-energy emission is absorbed by soft photons
in the source, which leads to a limit  $\Gamma\le 30$ on the bulk Lorentz factor \citep{2016ApJ...825...47P}.  Collapsing blue supergiants are attractive for the relatively long freefall times of their envelopes, allowing accretion to power a long-lasting central engine.  The simulation by \cite{Perna2018ApJ}  shows that the jet can emerge out of the extended envelope, and the resulting light curves resemble those observed in ULGRBs. In the simulation, however, the internal energy density was manually input to  enable
acceleration to an asymptotic Lorentz factor $\Gamma=300$, and the real baryon loading in this model still needs further investigation. 
Our finding of $\Gamma\ge 300$ for GRB 220627A may provide a new defining properties for ULGRBs and argues that only those models involving low baryon loading are feasible. Detailed numerical simulations of these models are needed to study how large the bulk Lorentz factors can be achieved \citep{Perna2018ApJ}. 

In summary, we report the detection of  GeV emission from an extremely long-duration GRB (GRB 220627A) with Fermi-LAT. The prompt emission of GRB 220627A  detected by Fermi-GBM has two episodes with a long quiescent phase inbetween. As the two episodes show a quite similar temporal shape and spectrum, GRB 220627A has been speculated to be a gravitational lensed GRB. Our analysis of the GeV emission, however, shows that the two episodes have quite a large difference in the flux ratio between the GeV emission and keV/MeV emission. This spectral difference  strongly argues against the lensed GRB scenario and we  conclude that GRB 220627A is an intrinsically ultralong GRB. The GeV emission exhibits a short spike coincident with the keV/MeV emission, suggesting that the GeV emission is of internal origin as the prompt keV/MeV emission. Detection of GeV photons also implies that the bulk Lorentz factors of the emitting region should be  $\Gamma\ge 300$. If this property applies to the class of ULGRBs, this will provide an important clue to distinguish  different models for ULGRBs.

\begin{acknowledgments}

We thank the anonymous referee for valuable suggestions. This work is supported by the NSFC grants Nos. 12121003, U2031105 and 12203022, the National Key R\&D program of China under grant No. 2018YFA0404203, China Manned Spaced Project (CMS-CSST-2021-B11), and the Natural Science Foundation of Jiangsu Province grant BK20220757.

\end{acknowledgments}

\newpage

\bibliography{reference}{}

\begin{thebibliography}{}
\expandafter\ifx\csname natexlab\endcsname\relax\def\natexlab#1{#1}\fi
\providecommand{\url}[1]{\href{#1}{#1}}
\providecommand{\dodoi}[1]{doi:~\href{http://doi.org/#1}{\nolinkurl{#1}}}
\providecommand{\doeprint}[1]{\href{http://ascl.net/#1}{\nolinkurl{http://ascl.net/#1}}}
\providecommand{\doarXiv}[1]{\href{https://arxiv.org/abs/#1}{\nolinkurl{https://arxiv.org/abs/#1}}}

\bibitem[{{Abbott} {et~al.}(2017){Abbott}, {Abbott}, {Abbott}, {Acernese},
  {Ackley}, {Adams}, {Adams}, {Addesso}, {Adhikari}, {Adya}, {Affeldt},
  {Afrough}, {Agarwal}, {Agathos}, {Agatsuma}, {Aggarwal}, {Aguiar}, {Aiello},
  {Ain}, {Ajith}, {Allen}, {Allen}, {Allocca}, {Altin}, {Amato}, {Ananyeva},
  {Anderson}, {Anderson}, {Angelova}, {Antier}, {Appert}, {Arai}, {Araya},
  {Areeda}, {Arnaud}, {Arun}, {Ascenzi}, {Ashton}, {Ast}, {Aston}, {Astone},
  {Atallah}, {Aufmuth}, {Aulbert}, {AultONeal}, {Austin}, {Avila-Alvarez},
  {Babak}, {Bacon}, {Bader}, {Bae}, {Baker}, {Baldaccini}, {Ballardin},
  {Ballmer}, {Banagiri}, {Barayoga}, {Barclay}, {Barish}, {Barker}, {Barkett},
  {Barone}, {Barr}, {Barsotti}, {Barsuglia}, {Barta}, {Barthelmy}, {Bartlett},
  {Bartos}, {Bassiri}, {Basti}, {Batch}, {Bawaj}, {Bayley}, {Bazzan},
  {B{\'e}csy}, {Beer}, {Bejger}, {Belahcene}, {Bell}, {Berger}, {Bergmann},
  {Bero}, {Berry}, {Bersanetti}, {Bertolini}, {Betzwieser}, {Bhagwat},
  {Bhandare}, {Bilenko}, {Billingsley}, {Billman}, {Birch}, {Birney},
  {Birnholtz}, {Biscans}, {Biscoveanu}, {Bisht}, {Bitossi}, {Biwer},
  {Bizouard}, {Blackburn}, {Blackman}, {Blair}, {Blair}, {Blair}, {Bloemen},
  {Bock}, {Bode}, {Boer}, {Bogaert}, {Bohe}, {Bondu}, {Bonilla}, {Bonnand},
  {Boom}, {Bork}, {Boschi}, {Bose}, {Bossie}, {Bouffanais}, {Bozzi},
  {Bradaschia}, {Brady}, {Branchesi}, {Brau}, {Briant}, {Brillet}, {Brinkmann},
  {Brisson}, {Brockill}, {Broida}, {Brooks}, {Brown}, {Brown}, {Brunett},
  {Buchanan}, {Buikema}, {Bulik}, {Bulten}, {Buonanno}, {Buskulic}, {Buy},
  {Byer}, {Cabero}, {Cadonati}, {Cagnoli}, {Cahillane}, {Calder{\'o}n
  Bustillo}, {Callister}, {Calloni}, {Camp}, {Canepa}, {Canizares}, {Cannon},
  {Cao}, {Cao}, {Capano}, {Capocasa}, {Carbognani}, {Caride}, {Carney},
  {Casanueva Diaz}, {Casentini}, {Caudill}, {Cavagli{\`a}}, {Cavalier},
  {Cavalieri}, {Cella}, {Cepeda}, {Cerd{\'a}-Dur{\'a}n}, {Cerretani},
  {Cesarini}, {Chamberlin}, {Chan}, {Chao}, {Charlton}, {Chase},
  {Chassande-Mottin}, {Chatterjee}, {Chatziioannou}, {Cheeseboro}, {Chen},
  {Chen}, {Chen}, {Cheng}, {Chia}, {Chincarini}, {Chiummo}, {Chmiel}, {Cho},
  {Cho}, {Chow}, {Christensen}, {Chu}, {Chua}, {Chua}, {Chung}, {Chung},
  {Ciani}, {Ciolfi}, {Cirelli}, {Cirone}, {Clara}, {Clark}, {Clearwater},
  {Cleva}, {Cocchieri}, {Coccia}, {Cohadon}, {Cohen}, {Colla}, {Collette},
  {Cominsky}, {Constancio}, {Conti}, {Cooper}, {Corban}, {Corbitt},
  {Cordero-Carri{\'o}n}, {Corley}, {Cornish}, {Corsi}, {Cortese}, {Costa},
  {Coughlin}, {Coughlin}, {Coulon}, {Countryman}, {Couvares}, {Covas}, {Cowan},
  {Coward}, {Cowart}, {Coyne}, {Coyne}, {Creighton}, {Creighton}, {Cripe},
  {Crowder}, {Cullen}, {Cumming}, {Cunningham}, {Cuoco}, {Dal Canton},
  {D{\'a}lya}, {Danilishin}, {D'Antonio}, {Danzmann}, {Dasgupta}, {Da Silva
  Costa}, {Dattilo}, {Dave}, {Davier}, {Davis}, {Daw}, {Day}, {De}, {DeBra},
  {Degallaix}, {De Laurentis}, {Del{\'e}glise}, {Del Pozzo}, {Demos}, {Denker},
  {Dent}, {De Pietri}, {Dergachev}, {De Rosa}, {DeRosa}, {De Rossi}, {DeSalvo},
  {de Varona}, {Devenson}, {Dhurandhar}, {D{\'\i}az}, {Di Fiore}, {Di
  Giovanni}, {Di Girolamo}, {Di Lieto}, {Di Pace}, {Di Palma}, {Di Renzo},
  {Doctor}, {Dolique}, {Donovan}, {Dooley}, {Doravari}, {Dorrington},
  {Douglas}, {Dovale {\'A}lvarez}, {Downes}, {Drago}, {Dreissigacker},
  {Driggers}, {Du}, {Ducrot}, {Dupej}, {Dwyer}, {Edo}, {Edwards}, {Effler},
  {Ehrens}, {Eichholz}, {Eikenberry}, {Eisenstein}, {Essick}, {Estevez},
  {Etienne}, {Etzel}, {Evans}, {Evans}, {Factourovich}, {Fafone}, {Fair},
  {Fairhurst}, {Fan}, {Farinon}, {Farr}, {Farr}, {Fauchon-Jones}, {Favata},
  {Fays}, {Fee}, {Fehrmann}, {Feicht}, {Fejer}, {Fernandez-Galiana},
  {Ferrante}, {Ferreira}, {Ferrini}, {Fidecaro}, {Finstad}, {Fiori},
  {Fiorucci}, {Fishbach}, {Fisher}, {Fitz-Axen}, {Flaminio}, {Fletcher},
  {Fong}, {Font}, {Forsyth}, {Forsyth}, {Fournier}, {Frasca}, {Frasconi},
  {Frei}, {Freise}, {Frey}, {Frey}, {Fries}, {Fritschel}, {Frolov}, {Fulda},
  {Fyffe}, {Gabbard}, {Gadre}, {Gaebel}, {Gair}, {Gammaitoni}, {Ganija},
  {Gaonkar}, {Garcia-Quiros}, {Garufi}, {Gateley}, {Gaudio}, {Gaur},
  {Gayathri}, {Gehrels}, {Gemme}, {Genin}, {Gennai}, {George}, {George},
  {Gergely}, {Germain}, {Ghonge}, {Ghosh}, {Ghosh}, {Ghosh}, {Giaime},
  {Giardina}, {Giazotto}, {Gill}, {Glover}, {Goetz}, {Goetz}, {Gomes},
  {Goncharov}, {Gonz{\'a}lez}, {Gonzalez Castro}, {Gopakumar}, {Gorodetsky},
  {Gossan}, {Gosselin}, {Gouaty}, {Grado}, {Graef}, {Granata}, {Grant}, {Gras},
  {Gray}, {Greco}, {Green}, {Gretarsson}, {Griswold}, {Groot}, {Grote},
  {Grunewald}, {Gruning}, {Guidi}, {Guo}, {Gupta}, {Gupta}, {Gushwa},
  {Gustafson}, {Gustafson}, {Halim}, {Hall}, {Hall}, {Hamilton}, {Hammond},
  {Haney}, {Hanke}, {Hanks}, {Hanna}, {Hannam}, {Hannuksela}, {Hanson},
  {Hardwick}, {Harms}, {Harry}, {Harry}, {Hart}, {Haster}, {Haughian}, {Healy},
  {Heidmann}, {Heintze}, {Heitmann}, {Hello}, {Hemming}, {Hendry}, {Heng},
  {Hennig}, {Heptonstall}, {Heurs}, {Hild}, {Hinderer}, {Hoak}, {Hofman},
  {Holt}, {Holz}, {Hopkins}, {Horst}, {Hough}, {Houston}, {Howell}, {Hreibi},
  {Hu}, {Huerta}, {Huet}, {Hughey}, {Husa}, {Huttner}, {Huynh-Dinh}, {Indik},
  {Inta}, {Intini}, {Isa}, {Isac}, {Isi}, {Iyer}, {Izumi}, {Jacqmin}, {Jani},
  {Jaranowski}, {Jawahar}, {Jim{\'e}nez-Forteza}, {Johnson}, {Jones}, {Jones},
  {Jonker}, {Ju}, {Junker}, {Kalaghatgi}, {Kalogera}, {Kamai}, {Kandhasamy},
  {Kang}, {Kanner}, {Kapadia}, {Karki}, {Karvinen}, {Kasprzack}, {Katolik},
  {Katsavounidis}, {Katzman}, {Kaufer}, {Kawabe}, {K{\'e}f{\'e}lian}, {Keitel},
  {Kemball}, {Kennedy}, {Kent}, {Key}, {Khalili}, {Khan}, {Khan}, {Khan},
  {Khazanov}, {Kijbunchoo}, {Kim}, {Kim}, {Kim}, {Kim}, {Kim}, {Kim},
  {Kimbrell}, {King}, {King}, {Kinley-Hanlon}, {Kirchhoff}, {Kissel},
  {Kleybolte}, {Klimenko}, {Knowles}, {Koch}, {Koehlenbeck}, {Koley},
  {Kondrashov}, {Kontos}, {Korobko}, {Korth}, {Kowalska}, {Kozak},
  {Kr{\"a}mer}, {Kringel}, {Krishnan}, {Kr{\'o}lak}, {Kuehn}, {Kumar}, {Kumar},
  {Kumar}, {Kuo}, {Kutynia}, {Kwang}, {Lackey}, {Lai}, {Landry}, {Lang},
  {Lange}, {Lantz}, {Lanza}, {Larson}, {Lartaux-Vollard}, {Lasky}, {Laxen},
  {Lazzarini}, {Lazzaro}, {Leaci}, {Leavey}, {Lee}, {Lee}, {Lee}, {Lee}, {Lee},
  {Lehmann}, {Lenon}, {Leonardi}, {Leroy}, {Letendre}, {Levin}, {Li}, {Linker},
  {Littenberg}, {Liu}, {Lo}, {Lockerbie}, {London}, {Lord}, {Lorenzini},
  {Loriette}, {Lormand}, {Losurdo}, {Lough}, {Lousto}, {Lovelace}, {L{\"u}ck},
  {Lumaca}, {Lundgren}, {Lynch}, {Ma}, {Macas}, {Macfoy}, {Machenschalk},
  {MacInnis}, {Macleod}, {Maga{\~n}a Hernandez}, {Maga{\~n}a-Sandoval},
  {Maga{\~n}a Zertuche}, {Magee}, {Majorana}, {Maksimovic}, {Man}, {Mandic},
  {Mangano}, {Mansell}, {Manske}, {Mantovani}, {Marchesoni}, {Marion},
  {M{\'a}rka}, {M{\'a}rka}, {Markakis}, {Markosyan}, {Markowitz}, {Maros},
  {Marquina}, {Marsh}, {Martelli}, {Martellini}, {Martin}, {Martin},
  {Martynov}, {Mason}, {Massera}, {Masserot}, {Massinger}, {Masso-Reid},
  {Mastrogiovanni}, {Matas}, {Matichard}, {Matone}, {Mavalvala}, {Mazumder},
  {McCarthy}, {McClelland}, {McCormick}, {McCuller}, {McGuire}, {McIntyre},
  {McIver}, {McManus}, {McNeill}, {McRae}, {McWilliams}, {Meacher}, {Meadors},
  {Mehmet}, {Meidam}, {Mejuto-Villa}, {Melatos}, {Mendell}, {Mercer}, {Merilh},
  {Merzougui}, {Meshkov}, {Messenger}, {Messick}, {Metzdorff}, {Meyers},
  {Miao}, {Michel}, {Middleton}, {Mikhailov}, {Milano}, {Miller}, {Miller},
  {Miller}, {Millhouse}, {Milovich-Goff}, {Minazzoli}, {Minenkov}, {Ming},
  {Mishra}, {Mitra}, {Mitrofanov}, {Mitselmakher}, {Mittleman}, {Moffa},
  {Moggi}, {Mogushi}, {Mohan}, {Mohapatra}, {Montani}, {Moore}, {Moraru},
  {Moreno}, {Morriss}, {Mours}, {Mow-Lowry}, {Mueller}, {Muir}, {Mukherjee},
  {Mukherjee}, {Mukherjee}, {Mukund}, {Mullavey}, {Munch}, {Mu{\~n}iz},
  {Muratore}, {Murray}, {Napier}, {Nardecchia}, {Naticchioni}, {Nayak},
  {Neilson}, {Nelemans}, {Nelson}, {Nery}, {Neunzert}, {Nevin}, {Newport},
  {Newton}, {Ng}, {Nguyen}, {Nguyen}, {Nichols}, {Nielsen}, {Nissanke}, {Nitz},
  {Noack}, {Nocera}, {Nolting}, {North}, {Nuttall}, {Oberling}, {O'Dea},
  {Ogin}, {Oh}, {Oh}, {Ohme}, {Okada}, {Oliver}, {Oppermann}, {Oram},
  {O'Reilly}, {Ormiston}, {Ortega}, {O'Shaughnessy}, {Ossokine}, {Ottaway},
  {Overmier}, {Owen}, {Pace}, {Page}, {Page}, {Pai}, {Pai}, {Palamos},
  {Palashov}, {Palomba}, {Pal-Singh}, {Pan}, {Pan}, {Pang}, {Pang}, {Pankow},
  {Pannarale}, {Pant}, {Paoletti}, {Paoli}, {Papa}, {Parida}, {Parker},
  {Pascucci}, {Pasqualetti}, {Passaquieti}, {Passuello}, {Patil}, {Patricelli},
  {Pearlstone}, {Pedraza}, {Pedurand}, {Pekowsky}, {Pele}, {Penn}, {Perez},
  {Perreca}, {Perri}, {Pfeiffer}, {Phelps}, {Piccinni}, {Pichot},
  {Piergiovanni}, {Pierro}, {Pillant}, {Pinard}, {Pinto}, {Pirello}, {Pitkin},
  {Poe}, {Poggiani}, {Popolizio}, {Porter}, {Post}, {Powell}, {Prasad},
  {Pratt}, {Pratten}, {Predoi}, {Prestegard}, {Price}, {Prijatelj}, {Principe},
  {Privitera}, {Prodi}, {Prokhorov}, {Puncken}, {Punturo}, {Puppo},
  {P{\"u}rrer}, {Qi}, {Quetschke}, {Quintero}, {Quitzow-James}, {Raab},
  {Rabeling}, {Radkins}, {Raffai}, {Raja}, {Rajan}, {Rajbhandari}, {Rakhmanov},
  {Ramirez}, {Ramos-Buades}, {Rapagnani}, {Raymond}, {Razzano}, {Read},
  {Regimbau}, {Rei}, {Reid}, {Reitze}, {Ren}, {Reyes}, {Ricci}, {Ricker},
  {Rieger}, {Riles}, {Rizzo}, {Robertson}, {Robie}, {Robinet}, {Rocchi},
  {Rolland}, {Rollins}, {Roma}, {Romano}, {Romel}, {Romie}, {Rosi{\'n}ska},
  {Ross}, {Rowan}, {R{\"u}diger}, {Ruggi}, {Rutins}, {Ryan}, {Sachdev},
  {Sadecki}, {Sadeghian}, {Sakellariadou}, {Salconi}, {Saleem}, {Salemi},
  {Samajdar}, {Sammut}, {Sampson}, {Sanchez}, {Sanchez}, {Sanchis-Gual},
  {Sandberg}, {Sanders}, {Sassolas}, {Sathyaprakash}, {Saulson}, {Sauter},
  {Savage}, {Sawadsky}, {Schale}, {Scheel}, {Scheuer}, {Schmidt}, {Schmidt},
  {Schnabel}, {Schofield}, {Sch{\"o}nbeck}, {Schreiber}, {Schuette}, {Schulte},
  {Schutz}, {Schwalbe}, {Scott}, {Scott}, {Seidel}, {Sellers}, {Sengupta},
  {Sentenac}, {Sequino}, {Sergeev}, {Shaddock}, {Shaffer}, {Shah}, {Shahriar},
  {Shaner}, {Shao}, {Shapiro}, {Shawhan}, {Sheperd}, {Shoemaker}, {Shoemaker},
  {Siellez}, {Siemens}, {Sieniawska}, {Sigg}, {Silva}, {Singer}, {Singh},
  {Singhal}, {Sintes}, {Slagmolen}, {Smith}, {Smith}, {Smith}, {Somala}, {Son},
  {Sonnenberg}, {Sorazu}, {Sorrentino}, {Souradeep}, {Spencer}, {Srivastava},
  {Staats}, {Staley}, {Steinke}, {Steinlechner}, {Steinlechner}, {Steinmeyer},
  {Stevenson}, {Stone}, {Stops}, {Strain}, {Stratta}, {Strigin}, {Strunk},
  {Sturani}, {Stuver}, {Summerscales}, {Sun}, {Sunil}, {Suresh}, {Sutton},
  {Swinkels}, {Szczepa{\'n}czyk}, {Tacca}, {Tait}, {Talbot}, {Talukder},
  {Tanner}, {T{\'a}pai}, {Taracchini}, {Tasson}, {Taylor}, {Taylor}, {Tewari},
  {Theeg}, {Thies}, {Thomas}, {Thomas}, {Thomas}, {Thorne}, {Thorne}, {Thrane},
  {Tiwari}, {Tiwari}, {Tokmakov}, {Toland}, {Tonelli}, {Tornasi},
  {Torres-Forn{\'e}}, {Torrie}, {T{\"o}yr{\"a}}, {Travasso}, {Traylor},
  {Trinastic}, {Tringali}, {Trozzo}, {Tsang}, {Tse}, {Tso}, {Tsukada}, {Tsuna},
  {Tuyenbayev}, {Ueno}, {Ugolini}, {Unnikrishnan}, {Urban}, {Usman},
  {Vahlbruch}, {Vajente}, {Valdes}, {van Bakel}, {van Beuzekom}, {van den
  Brand}, {Van Den Broeck}, {Vander-Hyde}, {van der Schaaf}, {van Heijningen},
  {van Veggel}, {Vardaro}, {Varma}, {Vass}, {Vas{\'u}th}, {Vecchio},
  {Vedovato}, {Veitch}, {Veitch}, {Venkateswara}, {Venugopalan}, {Verkindt},
  {Vetrano}, {Vicer{\'e}}, {Viets}, {Vinciguerra}, {Vine}, {Vinet}, {Vitale},
  {Vo}, {Vocca}, {Vorvick}, {Vyatchanin}, {Wade}, {Wade}, {Wade}, {Walet},
  {Walker}, {Wallace}, {Walsh}, {Wang}, {Wang}, {Wang}, {Wang}, {Wang}, {Ward},
  {Warner}, {Was}, {Watchi}, {Weaver}, {Wei}, {Weinert}, {Weinstein}, {Weiss},
  {Wen}, {Wessel}, {Wessels}, {Westerweck}, {Westphal}, {Wette}, {Whelan},
  {Whitcomb}, {Whiting}, {Whittle}, {Wilken}, {Williams}, {Williams},
  {Williamson}, {Willis}, {Willke}, {Wimmer}, {Winkler}, {Wipf}, {Wittel},
  {Woan}, {Woehler}, {Wofford}, {Wong}, {Worden}, {Wright}, {Wu}, {Wysocki},
  {Xiao}, {Yamamoto}, {Yancey}, {Yang}, {Yap}, {Yazback}, {Yu}, {Yu}, {Yvert},
  {Zadro{\.z}ny}, {Zanolin}, {Zelenova}, {Zendri}, {Zevin}, {Zhang}, {Zhang},
  {Zhang}, {Zhang}, {Zhao}, {Zhou}, {Zhou}, {Zhu}, {Zhu}, {Zimmerman},
  {Zucker}, {Zweizig}, {LIGO Scientific Collaboration}, {Virgo Collaboration},
  {Wilson-Hodge}, {Bissaldi}, {Blackburn}, {Briggs}, {Burns}, {Cleveland},
  {Connaughton}, {Gibby}, {Giles}, {Goldstein}, {Hamburg}, {Jenke}, {Hui},
  {Kippen}, {Kocevski}, {McBreen}, {Meegan}, {Paciesas}, {Poolakkil}, {Preece},
  {Racusin}, {Roberts}, {Stanbro}, {Veres}, {von Kienlin}, {GBM}, {Savchenko},
  {Ferrigno}, {Kuulkers}, {Bazzano}, {Bozzo}, {Brandt}, {Chenevez},
  {Courvoisier}, {Diehl}, {Domingo}, {Hanlon}, {Jourdain}, {Laurent}, {Lebrun},
  {Lutovinov}, {Martin-Carrillo}, {Mereghetti}, {Natalucci}, {Rodi}, {Roques},
  {Sunyaev}, {Ubertini}, {INTEGRAL}, {Aartsen}, {Ackermann}, {Adams},
  {Aguilar}, {Ahlers}, {Ahrens}, {Samarai}, {Altmann}, {Andeen}, {Anderson},
  {Ansseau}, {Anton}, {Arg{\"u}elles}, {Auffenberg}, {Axani}, {Bagherpour},
  {Bai}, {Barron}, {Barwick}, {Baum}, {Bay}, {Beatty}, {Becker Tjus},
  {Bernardini}, {Besson}, {Binder}, {Bindig}, {Blaufuss}, {Blot}, {Bohm},
  {B{\"o}rner}, {Bos}, {Bose}, {B{\"o}ser}, {Botner}, {Bourbeau}, {Bourbeau},
  {Bradascio}, {Braun}, {Brayeur}, {Brenzke}, {Bretz}, {Bron},
  {Brostean-Kaiser}, {Burgman}, {Carver}, {Casey}, {Casier}, {Cheung},
  {Chirkin}, {Christov}, {Clark}, {Classen}, {Coenders}, {Collin}, {Conrad},
  {Cowen}, {Cross}, {Day}, {de Andr{\'e}}, {De Clercq}, {DeLaunay},
  {Dembinski}, {De Ridder}, {Desiati}, {de Vries}, {de Wasseige}, {de With},
  {DeYoung}, {D{\'\i}az-V{\'e}lez}, {di Lorenzo}, {Dujmovic}, {Dumm},
  {Dunkman}, {Dvorak}, {Eberhardt}, {Ehrhardt}, {Eichmann}, {Eller}, {Evenson},
  {Fahey}, {Fazely}, {Felde}, {Filimonov}, {Finley}, {Flis}, {Franckowiak},
  {Friedman}, {Fuchs}, {Gaisser}, {Gallagher}, {Gerhardt}, {Ghorbani}, {Giang},
  {Glauch}, {Gl{\"u}senkamp}, {Goldschmidt}, {Gonzalez}, {Grant}, {Griffith},
  {Haack}, {Hallgren}, {Halzen}, {Hanson}, {Hebecker}, {Heereman}, {Helbing},
  {Hellauer}, {Hickford}, {Hignight}, {Hill}, {Hoffman}, {Hoffmann},
  {Hokanson-Fasig}, {Hoshina}, {Huang}, {Huber}, {Hultqvist}, {H{\"u}nnefeld},
  {In}, {Ishihara}, {Jacobi}, {Japaridze}, {Jeong}, {Jero}, {Jones},
  {Kalaczynski}, {Kang}, {Kappes}, {Karg}, {Karle}, {Kauer}, {Keivani},
  {Kelley}, {Kheirandish}, {Kim}, {Kim}, {Kintscher}, {Kiryluk}, {Kittler},
  {Klein}, {Kohnen}, {Koirala}, {Kolanoski}, {K{\"o}pke}, {Kopper}, {Kopper},
  {Koschinsky}, {Koskinen}, {Kowalski}, {Krings}, {Kroll}, {Kr{\"u}ckl},
  {Kunnen}, {Kunwar}, {Kurahashi}, {Kuwabara}, {Kyriacou}, {Labare},
  {Lanfranchi}, {Larson}, {Lauber}, {Lesiak-Bzdak}, {Leuermann}, {Liu}, {Lu},
  {L{\"u}nemann}, {Luszczak}, {Madsen}, {Maggi}, {Mahn}, {Mancina}, {Maruyama},
  {Mase}, {Maunu}, {McNally}, {Meagher}, {Medici}, {Meier}, {Menne}, {Merino},
  {Meures}, {Miarecki}, {Micallef}, {Moment{\'e}}, {Montaruli}, {Moore},
  {Moulai}, {Nahnhauer}, {Nakarmi}, {Naumann}, {Neer}, {Niederhausen},
  {Nowicki}, {Nygren}, {Obertacke Pollmann}, {Olivas}, {O'Murchadha},
  {Palczewski}, {Pandya}, {Pankova}, {Peiffer}, {Pepper}, {P{\'e}rez de los
  Heros}, {Pieloth}, {Pinat}, {Price}, {Przybylski}, {Raab}, {R{\"a}del},
  {Rameez}, {Rawlins}, {Rea}, {Reimann}, {Relethford}, {Relich}, {Resconi},
  {Rhode}, {Richman}, {Robertson}, {Rongen}, {Rott}, {Ruhe}, {Ryckbosch},
  {Rysewyk}, {S{\"a}lzer}, {Sanchez Herrera}, {Sandrock}, {Sandroos},
  {Santander}, {Sarkar}, {Sarkar}, {Satalecka}, {Schlunder}, {Schmidt},
  {Schneider}, {Schoenen}, {Sch{\"o}neberg}, {Schumacher}, {Seckel},
  {Seunarine}, {Soedingrekso}, {Soldin}, {Song}, {Spiczak}, {Spiering},
  {Stachurska}, {Stamatikos}, {Stanev}, {Stasik}, {Stettner}, {Steuer},
  {Stezelberger}, {Stokstad}, {St{\"o}ssl}, {Strotjohann}, {Stuttard},
  {Sullivan}, {Sutherland}, {Taboada}, {Tatar}, {Tenholt}, {Ter-Antonyan},
  {Terliuk}, {Te{\v{s}}i{\'c}}, {Tilav}, {Toale}, {Tobin}, {Toscano}, {Tosi},
  {Tselengidou}, {Tung}, {Turcati}, {Turley}, {Ty}, {Unger}, {Usner},
  {Vandenbroucke}, {Van Driessche}, {van Eijndhoven}, {Vanheule}, {van Santen},
  {Vehring}, {Vogel}, {Vraeghe}, {Walck}, {Wallace}, {Wallraff}, {Wandler},
  {Wandkowsky}, {Waza}, {Weaver}, {Weiss}, {Wendt}, {Werthebach}, {Whelan},
  {Wiebe}, {Wiebusch}, {Wille}, {Williams}, {Wills}, {Wolf}, {Wood}, {Woolsey},
  {Woschnagg}, {Xu}, {Xu}, {Xu}, {Yanez}, {Yodh}, {Yoshida}, {Yuan}, {Zoll},
  {IceCube Collaboration}, {Balasubramanian}, {Mate}, {Bhalerao},
  {Bhattacharya}, {Vibhute}, {Dewangan}, {Rao}, {Vadawale}, {AstroSat Cadmium
  Zinc Telluride Imager Team}, {Svinkin}, {Hurley}, {Aptekar}, {Frederiks},
  {Golenetskii}, {Kozlova}, {Lysenko}, {Oleynik}, {Tsvetkova}, {Ulanov},
  {Cline}, {IPN Collaboration}, {Li}, {Xiong}, {Zhang}, {Lu}, {Song}, {Cao},
  {Chang}, {Chen}, {Chen}, {Chen}, {Chen}, {Chen}, {Chen}, {Cui}, {Cui},
  {Deng}, {Dong}, {Du}, {Fu}, {Gao}, {Gao}, {Gao}, {Ge}, {Gu}, {Guan}, {Guo},
  {Han}, {Hu}, {Huang}, {Huo}, {Jia}, {Jiang}, {Jiang}, {Jin}, {Jin}, {Li},
  {Li}, {Li}, {Li}, {Li}, {Li}, {Li}, {Li}, {Li}, {Li}, {Li}, {Liang}, {Liao},
  {Liu}, {Liu}, {Liu}, {Liu}, {Liu}, {Liu}, {Liu}, {Lu}, {Lu}, {Luo}, {Ma},
  {Meng}, {Nang}, {Nie}, {Ou}, {Qu}, {Sai}, {Sun}, {Tan}, {Tao}, {Tao}, {Tuo},
  {Wang}, {Wang}, {Wang}, {Wang}, {Wang}, {Wen}, {Wu}, {Wu}, {Xiao}, {Xu},
  {Xu}, {Yan}, {Yang}, {Yang}, {Yang}, {Zhang}, {Zhang}, {Zhang}, {Zhang},
  {Zhang}, {Zhang}, {Zhang}, {Zhang}, {Zhang}, {Zhang}, {Zhang}, {Zhang},
  {Zhang}, {Zhang}, {Zhang}, {Zhang}, {Zhang}, {Zhang}, {Zhao}, {Zhao}, {Zhao},
  {Zheng}, {Zhu}, {Zhu}, {Zou}, {Insight-HXMT Collaboration}, {Albert},
  {Andr{\'e}}, {Anghinolfi}, {Ardid}, {Aubert}, {Aublin}, {Avgitas}, {Baret},
  {Barrios-Mart{\'\i}}, {Basa}, {Belhorma}, {Bertin}, {Biagi}, {Bormuth},
  {Bourret}, {Bouwhuis}, {Br{\^a}nza{\c{s}}}, {Bruijn}, {Brunner}, {Busto},
  {Capone}, {Caramete}, {Carr}, {Celli}, {Cherkaoui El Moursli}, {Chiarusi},
  {Circella}, {Coelho}, {Coleiro}, {Coniglione}, {Costantini}, {Coyle},
  {Creusot}, {D{\'\i}az}, {Deschamps}, {De Bonis}, {Distefano}, {Di Palma},
  {Domi}, {Donzaud}, {Dornic}, {Drouhin}, {Eberl}, {El Bojaddaini}, {El
  Khayati}, {Els{\"a}sser}, {Enzenh{\"o}fer}, {Ettahiri}, {Fassi}, {Felis},
  {Fusco}, {Gay}, {Giordano}, {Glotin}, {Gr{\'e}goire}, {Ruiz}, {Graf},
  {Hallmann}, {van Haren}, {Heijboer}, {Hello}, {Hern{\'a}ndez-Rey},
  {H{\"o}ssl}, {Hofest{\"a}dt}, {Hugon}, {Illuminati}, {James}, {de Jong},
  {Jongen}, {Kadler}, {Kalekin}, {Katz}, {Kiessling}, {Kouchner}, {Kreter},
  {Kreykenbohm}, {Kulikovskiy}, {Lachaud}, {Lahmann}, {Lef{\`e}vre}, {Leonora},
  {Lotze}, {Loucatos}, {Marcelin}, {Margiotta}, {Marinelli},
  {Mart{\'\i}nez-Mora}, {Mele}, {Melis}, {Michael}, {Migliozzi}, {Moussa},
  {Navas}, {Nezri}, {Organokov}, {P{\u{a}}v{\u{a}}la{\c{s}}}, {Pellegrino},
  {Perrina}, {Piattelli}, {Popa}, {Pradier}, {Quinn}, {Racca}, {Riccobene},
  {S{\'a}nchez-Losa}, {Salda{\~n}a}, {Salvadori}, {Samtleben}, {Sanguineti},
  {Sapienza}, {Sieger}, {Spurio}, {Stolarczyk}, {Taiuti}, {Tayalati},
  {Trovato}, {Turpin}, {T{\"o}nnis}, {Vallage}, {Van Elewyck}, {Versari},
  {Vivolo}, {Vizzoca}, {Wilms}, {Zornoza}, {Z{\'u}{\~n}iga}, {ANTARES
  Collaboration}, {Beardmore}, {Breeveld}, {Burrows}, {Cenko}, {Cusumano},
  {D'A{\`\i}}, {de Pasquale}, {Emery}, {Evans}, {Giommi}, {Gronwall}, {Kennea},
  {Krimm}, {Kuin}, {Lien}, {Marshall}, {Melandri}, {Nousek}, {Oates},
  {Osborne}, {Pagani}, {Page}, {Palmer}, {Perri}, {Siegel}, {Sbarufatti},
  {Tagliaferri}, {Tohuvavohu}, {Swift Collaboration}, {Tavani}, {Verrecchia},
  {Bulgarelli}, {Evangelista}, {Pacciani}, {Feroci}, {Pittori}, {Giuliani},
  {Del Monte}, {Donnarumma}, {Argan}, {Trois}, {Ursi}, {Cardillo}, {Piano},
  {Longo}, {Lucarelli}, {Munar-Adrover}, {Fuschino}, {Labanti}, {Marisaldi},
  {Minervini}, {Fioretti}, {Parmiggiani}, {Gianotti}, {Trifoglio}, {Di Persio},
  {Antonelli}, {Barbiellini}, {Caraveo}, {Cattaneo}, {Costa}, {Colafrancesco},
  {D'Amico}, {Ferrari}, {Morselli}, {Paoletti}, {Picozza}, {Pilia}, {Rappoldi},
  {Soffitta}, {Vercellone}, {AGILE Team}, {Foley}, {Coulter}, {Kilpatrick},
  {Drout}, {Piro}, {Shappee}, {Siebert}, {Simon}, {Ulloa}, {Kasen}, {Madore},
  {Murguia-Berthier}, {Pan}, {Prochaska}, {Ramirez-Ruiz}, {Rest},
  {Rojas-Bravo}, {1M2H Team}, {Berger}, {Soares-Santos}, {Annis}, {Alexander},
  {Allam}, {Balbinot}, {Blanchard}, {Brout}, {Butler}, {Chornock}, {Cook},
  {Cowperthwaite}, {Diehl}, {Drlica-Wagner}, {Drout}, {Durret}, {Eftekhari},
  {Finley}, {Fong}, {Frieman}, {Fryer}, {Garc{\'\i}a-Bellido}, {Gruendl},
  {Hartley}, {Herner}, {Kessler}, {Lin}, {Lopes}, {Louren{\c{c}}o}, {Margutti},
  {Marshall}, {Matheson}, {Medina}, {Metzger}, {Mu{\~n}oz}, {Muir}, {Nicholl},
  {Nugent}, {Palmese}, {Paz-Chinch{\'o}n}, {Quataert}, {Sako}, {Sauseda},
  {Schlegel}, {Scolnic}, {Secco}, {Smith}, {Sobreira}, {Villar}, {Vivas},
  {Wester}, {Williams}, {Yanny}, {Zenteno}, {Zhang}, {Abbott}, {Banerji},
  {Bechtol}, {Benoit-L{\'e}vy}, {Bertin}, {Brooks}, {Buckley-Geer}, {Burke},
  {Capozzi}, {Carnero Rosell}, {Carrasco Kind}, {Castander}, {Crocce}, {Cunha},
  {D'Andrea}, {da Costa}, {Davis}, {DePoy}, {Desai}, {Dietrich}, {Eifler},
  {Fernandez}, {Flaugher}, {Fosalba}, {Gaztanaga}, {Gerdes}, {Giannantonio},
  {Goldstein}, {Gruen}, {Gschwend}, {Gutierrez}, {Honscheid}, {James},
  {Jeltema}, {Johnson}, {Johnson}, {Kent}, {Krause}, {Kron}, {Kuehn}, {Lahav},
  {Lima}, {Maia}, {March}, {Martini}, {McMahon}, {Menanteau}, {Miller},
  {Miquel}, {Mohr}, {Nichol}, {Ogando}, {Plazas}, {Romer}, {Roodman}, {Rykoff},
  {Sanchez}, {Scarpine}, {Schindler}, {Schubnell}, {Sevilla-Noarbe}, {Sheldon},
  {Smith}, {Smith}, {Stebbins}, {Suchyta}, {Swanson}, {Tarle}, {Thomas},
  {Troxel}, {Tucker}, {Vikram}, {Walker}, {Wechsler}, {Weller}, {Carlin},
  {Gill}, {Li}, {Marriner}, {Neilsen}, {Dark Energy Camera GW-EM
  Collaboration}, {DES Collaboration}, {Haislip}, {Kouprianov}, {Reichart},
  {Sand}, {Tartaglia}, {Valenti}, {Yang}, {DLT40 Collaboration}, {Benetti},
  {Brocato}, {Campana}, {Cappellaro}, {Covino}, {D'Avanzo}, {D'Elia}, {Getman},
  {Ghirlanda}, {Ghisellini}, {Limatola}, {Nicastro}, {Palazzi}, {Pian},
  {Piranomonte}, {Possenti}, {Rossi}, {Salafia}, {Tomasella}, {Amati},
  {Antonelli}, {Bernardini}, {Bufano}, {Capaccioli}, {Casella}, {Dadina}, {De
  Cesare}, {Di Paola}, {Giuffrida}, {Giunta}, {Israel}, {Lisi}, {Maiorano},
  {Mapelli}, {Masetti}, {Pescalli}, {Pulone}, {Salvaterra}, {Schipani},
  {Spera}, {Stamerra}, {Stella}, {Testa}, {Turatto}, {Vergani}, {Aresu},
  {Bachetti}, {Buffa}, {Burgay}, {Buttu}, {Caria}, {Carretti}, {Casasola},
  {Castangia}, {Carboni}, {Casu}, {Concu}, {Corongiu}, {Deiana}, {Egron},
  {Fara}, {Gaudiomonte}, {Gusai}, {Ladu}, {Loru}, {Leurini}, {Marongiu},
  {Melis}, {Melis}, {Migoni}, {Milia}, {Navarrini}, {Orlati}, {Ortu}, {Palmas},
  {Pellizzoni}, {Perrodin}, {Pisanu}, {Poppi}, {Righini}, {Saba}, {Serra},
  {Serrau}, {Stagni}, {Surcis}, {Vacca}, {Vargiu}, {Hunt}, {Jin}, {Klose},
  {Kouveliotou}, {Mazzali}, {M{\o}ller}, {Nava}, {Piran}, {Selsing}, {Vergani},
  {Wiersema}, {Toma}, {Higgins}, {Mundell}, {di Serego Alighieri}, {G{\'o}tz},
  {Gao}, {Gomboc}, {Kaper}, {Kobayashi}, {Kopac}, {Mao}, {Starling}, {Steele},
  {van der Horst}, {GRAWITA: GRAvitational Wave Inaf TeAm}, {Acero}, {Atwood},
  {Baldini}, {Barbiellini}, {Bastieri}, {Berenji}, {Bellazzini}, {Bissaldi},
  {Blandford}, {Bloom}, {Bonino}, {Bottacini}, {Bregeon}, {Buehler}, {Buson},
  {Cameron}, {Caputo}, {Caraveo}, {Cavazzuti}, {Chekhtman}, {Cheung}, {Chiang},
  {Ciprini}, {Cohen-Tanugi}, {Cominsky}, {Costantin}, {Cuoco}, {D'Ammando}, {de
  Palma}, {Digel}, {Di Lalla}, {Di Mauro}, {Di Venere}, {Dubois}, {Fegan},
  {Focke}, {Franckowiak}, {Fukazawa}, {Funk}, {Fusco}, {Gargano}, {Gasparrini},
  {Giglietto}, {Giordano}, {Giroletti}, {Glanzman}, {Green}, {Grondin},
  {Guillemot}, {Guiriec}, {Harding}, {Horan}, {J{\'o}hannesson}, {Kamae},
  {Kensei}, {Kuss}, {La Mura}, {Latronico}, {Lemoine-Goumard}, {Longo},
  {Loparco}, {Lovellette}, {Lubrano}, {Magill}, {Maldera}, {Manfreda},
  {Mazziotta}, {McEnery}, {Meyer}, {Michelson}, {Mirabal}, {Monzani},
  {Moretti}, {Morselli}, {Moskalenko}, {Negro}, {Nuss}, {Ojha}, {Omodei},
  {Orienti}, {Orlando}, {Palatiello}, {Paliya}, {Paneque}, {Pesce-Rollins},
  {Piron}, {Porter}, {Principe}, {Rain{\`o}}, {Rando}, {Razzano}, {Razzaque},
  {Reimer}, {Reimer}, {Reposeur}, {Rochester}, {Saz Parkinson}, {Sgr{\`o}},
  {Siskind}, {Spada}, {Spandre}, {Suson}, {Takahashi}, {Tanaka}, {Thayer},
  {Thayer}, {Thompson}, {Tibaldo}, {Torres}, {Torresi}, {Troja}, {Venters},
  {Vianello}, {Zaharijas}, {Fermi Large Area Telescope Collaboration},
  {Allison}, {Bannister}, {Dobie}, {Kaplan}, {Lenc}, {Lynch}, {Murphy},
  {Sadler}, {Australia Telescope Compact Array}, {Hotan}, {James}, {Oslowski},
  {Raja}, {Shannon}, {Whiting}, {Australian SKA Pathfinder}, {Arcavi},
  {Howell}, {McCully}, {Hosseinzadeh}, {Hiramatsu}, {Poznanski}, {Barnes},
  {Zaltzman}, {Vasylyev}, {Maoz}, {Las Cumbres Observatory Group}, {Cooke},
  {Bailes}, {Wolf}, {Deller}, {Lidman}, {Wang}, {Gendre}, {Andreoni}, {Ackley},
  {Pritchard}, {Bessell}, {Chang}, {M{\"o}ller}, {Onken}, {Scalzo},
  {Ridden-Harper}, {Sharp}, {Tucker}, {Farrell}, {Elmer}, {Johnston},
  {Venkatraman Krishnan}, {Keane}, {Green}, {Jameson}, {Hu}, {Ma}, {Sun}, {Wu},
  {Wang}, {Shang}, {Hu}, {Ashley}, {Yuan}, {Li}, {Tao}, {Zhu}, {Zhang},
  {Suntzeff}, {Zhou}, {Yang}, {Orange}, {Morris}, {Cucchiara}, {Giblin},
  {Klotz}, {Staff}, {Thierry}, {Schmidt}, {OzGrav}, {(Deeper}, {Wider},
  {program}, {AST3}, {CAASTRO Collaborations}, {Tanvir}, {Levan}, {Cano}, {de
  Ugarte-Postigo}, {Gonz{\'a}lez-Fern{\'a}ndez}, {Greiner}, {Hjorth}, {Irwin},
  {Kr{\"u}hler}, {Mandel}, {Milvang-Jensen}, {O'Brien}, {Rol}, {Rosetti},
  {Rosswog}, {Rowlinson}, {Steeghs}, {Th{\"o}ne}, {Ulaczyk}, {Watson}, {Bruun},
  {Cutter}, {Figuera Jaimes}, {Fujii}, {Fruchter}, {Gompertz}, {Jakobsson},
  {Hodosan}, {J{\`e}rgensen}, {Kangas}, {Kann}, {Rabus}, {Schr{\o}der},
  {Stanway}, {Wijers}, {VINROUGE Collaboration}, {Lipunov}, {Gorbovskoy},
  {Kornilov}, {Tyurina}, {Balanutsa}, {Kuznetsov}, {Vlasenko}, {Podesta},
  {Lopez}, {Podesta}, {Levato}, {Saffe}, {Mallamaci}, {Budnev}, {Gress},
  {Kuvshinov}, {Gorbunov}, {Vladimirov}, {Zimnukhov}, {Gabovich}, {Yurkov},
  {Sergienko}, {Rebolo}, {Serra-Ricart}, {Tlatov}, {Ishmuhametova}, {MASTER
  Collaboration}, {Abe}, {Aoki}, {Aoki}, {Asakura}, {Baar}, {Barway}, {Bond},
  {Doi}, {Finet}, {Fujiyoshi}, {Furusawa}, {Honda}, {Itoh}, {Kanda},
  {Kawabata}, {Kawabata}, {Kim}, {Koshida}, {Kuroda}, {Lee}, {Liu},
  {Matsubayashi}, {Miyazaki}, {Morihana}, {Morokuma}, {Motohara}, {Murata},
  {Nagai}, {Nagashima}, {Nagayama}, {Nakaoka}, {Nakata}, {Ohsawa}, {Ohshima},
  {Ohta}, {Okita}, {Saito}, {Saito}, {Sako}, {Sekiguchi}, {Sumi}, {Tajitsu},
  {Takahashi}, {Takayama}, {Tamura}, {Tanaka}, {Tanaka}, {Terai}, {Tominaga},
  {Tristram}, {Uemura}, {Utsumi}, {Yamaguchi}, {Yasuda}, {Yoshida}, {Zenko},
  {J-GEM}, {Adams}, {Anupama}, {Bally}, {Barway}, {Bellm}, {Blagorodnova},
  {Cannella}, {Chandra}, {Chatterjee}, {Clarke}, {Cobb}, {Cook}, {Copperwheat},
  {De}, {Emery}, {Feindt}, {Foster}, {Fox}, {Frail}, {Fremling}, {Frohmaier},
  {Garcia}, {Ghosh}, {Giacintucci}, {Goobar}, {Gottlieb}, {Grefenstette},
  {Hallinan}, {Harrison}, {Heida}, {Helou}, {Ho}, {Horesh}, {Hotokezaka}, {Ip},
  {Itoh}, {Jacobs}, {Jencson}, {Kasen}, {Kasliwal}, {Kassim}, {Kim}, {Kiran},
  {Kuin}, {Kulkarni}, {Kupfer}, {Lau}, {Madsen}, {Mazzali}, {Miller},
  {Miyasaka}, {Mooley}, {Myers}, {Nakar}, {Ngeow}, {Nugent}, {Ofek},
  {Palliyaguru}, {Pavana}, {Perley}, {Peters}, {Pike}, {Piran}, {Qi}, {Quimby},
  {Rana}, {Rosswog}, {Rusu}, {Sadler}, {Van Sistine}, {Sollerman}, {Xu}, {Yan},
  {Yatsu}, {Yu}, {Zhang}, {Zhao}, {GROWTH}, {JAGWAR}, {Caltech-NRAO},
  {TTU-NRAO}, {NuSTAR Collaborations}, {Chambers}, {Huber}, {Schultz},
  {Bulger}, {Flewelling}, {Magnier}, {Lowe}, {Wainscoat}, {Waters}, {Willman},
  {Pan-STARRS}, {Ebisawa}, {Hanyu}, {Harita}, {Hashimoto}, {Hidaka}, {Hori},
  {Ishikawa}, {Isobe}, {Iwakiri}, {Kawai}, {Kawai}, {Kawamuro}, {Kawase},
  {Kitaoka}, {Makishima}, {Matsuoka}, {Mihara}, {Morita}, {Morita}, {Nakahira},
  {Nakajima}, {Nakamura}, {Negoro}, {Oda}, {Sakamaki}, {Sasaki}, {Serino},
  {Shidatsu}, {Shimomukai}, {Sugawara}, {Sugita}, {Sugizaki}, {Tachibana},
  {Takao}, {Tanimoto}, {Tomida}, {Tsuboi}, {Tsunemi}, {Ueda}, {Ueno}, {Yamada},
  {Yamaoka}, {Yamauchi}, {Yatabe}, {Yoneyama}, {Yoshii}, {MAXI Team}, {Coward},
  {Crisp}, {Macpherson}, {Andreoni}, {Laugier}, {Noysena}, {Klotz}, {Gendre},
  {Thierry}, {Turpin}, {Consortium}, {Im}, {Choi}, {Kim}, {Yoon}, {Lim}, {Lee},
  {Lee}, {Kim}, {Ko}, {Joe}, {Kwon}, {Kim}, {Lim}, {Choi}, {KU Collaboration},
  {Fynbo}, {Malesani}, {Xu}, {Optical Telescope}, {Smartt}, {Jerkstrand},
  {Kankare}, {Sim}, {Fraser}, {Inserra}, {Maguire}, {Leloudas}, {Magee},
  {Shingles}, {Smith}, {Young}, {Kotak}, {Gal-Yam}, {Lyman}, {Homan},
  {Agliozzo}, {Anderson}, {Angus}, {Ashall}, {Barbarino}, {Bauer}, {Berton},
  {Botticella}, {Bulla}, {Cannizzaro}, {Cartier}, {Cikota}, {Clark}, {De Cia},
  {Della Valle}, {Dennefeld}, {Dessart}, {Dimitriadis}, {Elias-Rosa}, {Firth},
  {Fl{\"o}rs}, {Frohmaier}, {Galbany}, {Gonz{\'a}lez-Gait{\'a}n}, {Gromadzki},
  {Guti{\'e}rrez}, {Hamanowicz}, {Harmanen}, {Heintz}, {Hernandez}, {Hodgkin},
  {Hook}, {Izzo}, {James}, {Jonker}, {Kerzendorf}, {Kostrzewa-Rutkowska},
  {Kromer}, {Kuncarayakti}, {Lawrence}, {Manulis}, {Mattila}, {McBrien},
  {M{\"u}ller}, {Nordin}, {O'Neill}, {Onori}, {Palmerio}, {Pastorello},
  {Patat}, {Pignata}, {Podsiadlowski}, {Razza}, {Reynolds}, {Roy}, {Ruiter},
  {Rybicki}, {Salmon}, {Pumo}, {Prentice}, {Seitenzahl}, {Smith}, {Sollerman},
  {Sullivan}, {Szegedi}, {Taddia}, {Taubenberger}, {Terreran}, {Van Soelen},
  {Vos}, {Walton}, {Wright}, {Wyrzykowski}, {Yaron}, {pre=''(''>ePESSTO},
  {Chen}, {Kr{\"u}hler}, {Schady}, {Wiseman}, {Greiner}, {Rau}, {Schweyer},
  {Klose}, {Nicuesa Guelbenzu}, {GROND}, {Palliyaguru}, {Tech University},
  {Shara}, {Williams}, {Vaisanen}, {Potter}, {Romero Colmenero}, {Crawford},
  {Buckley}, {Mao}, {SALT Group}, {D{\'\i}az}, {Macri}, {Garc{\'\i}a Lambas},
  {Mendes de Oliveira}, {Nilo Castell{\'o}n}, {Ribeiro}, {S{\'a}nchez},
  {Schoenell}, {Abramo}, {Akras}, {Alcaniz}, {Artola}, {Beroiz}, {Bonoli},
  {Cabral}, {Camuccio}, {Chavushyan}, {Coelho}, {Colazo}, {Costa-Duarte},
  {Cuevas Larenas}, {Dom{\'\i}nguez Romero}, {Dultzin}, {Fern{\'a}ndez},
  {Garc{\'\i}a}, {Girardini}, {Gon{\c{c}}alves}, {Gon{\c{c}}alves}, {Gurovich},
  {Jim{\'e}nez-Teja}, {Kanaan}, {Lares}, {Lopes de Oliveira}, {L{\'o}pez-Cruz},
  {Melia}, {Molino}, {Padilla}, {Pe{\~n}uela}, {Placco}, {Qui{\~n}ones},
  {Ram{\'\i}rez Rivera}, {Renzi}, {Riguccini}, {R{\'\i}os-L{\'o}pez},
  {Rodriguez}, {Sampedro}, {Schneiter}, {Sodr{\'e}}, {Starck}, {Torres-Flores},
  {Tornatore}, {Zadro{\.z}ny}, {Castillo}, {TOROS: Transient Robotic
  Observatory of South Collaboration}, {Castro-Tirado}, {Tello}, {Hu}, {Zhang},
  {Cunniffe}, {Castell{\'o}n}, {Hiriart}, {Caballero-Garc{\'\i}a},
  {Jel{\'\i}nek}, {Kub{\'a}nek}, {P{\'e}rez del Pulgar}, {Park}, {Jeong},
  {Castro Cer{\'o}n}, {Pandey}, {Yock}, {Querel}, {Fan}, {Wang}, {BOOTES
  Collaboration}, {Beardsley}, {Brown}, {Crosse}, {Emrich}, {Franzen},
  {Gaensler}, {Horsley}, {Johnston-Hollitt}, {Kenney}, {Morales}, {Pallot},
  {Sokolowski}, {Steele}, {Tingay}, {Trott}, {Walker}, {Wayth}, {Williams},
  {Wu}, {Murchison Widefield Array}, {Yoshida}, {Sakamoto}, {Kawakubo},
  {Yamaoka}, {Takahashi}, {Asaoka}, {Ozawa}, {Torii}, {Shimizu}, {Tamura},
  {Ishizaki}, {Cherry}, {Ricciarini}, {Penacchioni}, {Marrocchesi}, {CALET
  Collaboration}, {Pozanenko}, {Volnova}, {Mazaeva}, {Minaev}, {Krugov},
  {Kusakin}, {Reva}, {Moskvitin}, {Rumyantsev}, {Inasaridze}, {Klunko},
  {Tungalag}, {Schmalz}, {Burhonov}, {IKI-GW Follow-up Collaboration},
  {Abdalla}, {Abramowski}, {Aharonian}, {Ait Benkhali}, {Ang{\"u}ner},
  {Arakawa}, {Arrieta}, {Aubert}, {Backes}, {Balzer}, {Barnard}, {Becherini},
  {Becker Tjus}, {Berge}, {Bernhard}, {Bernl{\"o}hr}, {Blackwell},
  {B{\"o}ttcher}, {Boisson}, {Bolmont}, {Bonnefoy}, {Bordas}, {Bregeon},
  {Brun}, {Brun}, {Bryan}, {B{\"u}chele}, {Bulik}, {Capasso}, {Caroff},
  {Carosi}, {Casanova}, {Cerruti}, {Chakraborty}, {Chaves}, {Chen},
  {Chevalier}, {Colafrancesco}, {Condon}, {Conrad}, {Davids}, {Decock}, {Deil},
  {Devin}, {deWilt}, {Dirson}, {Djannati-Ata{\"\i}}, {Donath}, {O'C. Drury},
  {Dutson}, {Dyks}, {Edwards}, {Egberts}, {Emery}, {Ernenwein}, {Eschbach},
  {Farnier}, {Fegan}, {Fernandes}, {Fiasson}, {Fontaine}, {Funk},
  {F{\"u}ssling}, {Gabici}, {Gallant}, {Garrigoux}, {Gat{\'e}}, {Giavitto},
  {Giebels}, {Glawion}, {Glicenstein}, {Gottschall}, {Grondin}, {Hahn},
  {Haupt}, {Hawkes}, {Heinzelmann}, {Henri}, {Hermann}, {Hinton}, {Hofmann},
  {Hoischen}, {Holch}, {Holler}, {Horns}, {Ivascenko}, {Iwasaki},
  {Jacholkowska}, {Jamrozy}, {Jankowsky}, {Jankowsky}, {Jingo}, {Jouvin},
  {Jung-Richardt}, {Kastendieck}, {Katarzy{\'n}ski}, {Katsuragawa},
  {Kerszberg}, {Khangulyan}, {Kh{\'e}lifi}, {King}, {Klepser}, {Klochkov},
  {Klu{\'z}niak}, {Komin}, {Kosack}, {Krakau}, {Kraus}, {Kr{\"u}ger}, {Laffon},
  {Lamanna}, {Lau}, {Lees}, {Lefaucheur}, {Lemi{\`e}re}, {Lemoine-Goumard},
  {Lenain}, {Leser}, {Lohse}, {Lorentz}, {Liu}, {Lypova}, {Malyshev},
  {Marandon}, {Marcowith}, {Mariaud}, {Marx}, {Maurin}, {Maxted}, {Mayer},
  {Meintjes}, {Meyer}, {Mitchell}, {Moderski}, {Mohamed}, {Mohrmann},
  {Mor{\r{a}}}, {Moulin}, {Murach}, {Nakashima}, {de Naurois}, {Ndiyavala},
  {Niederwanger}, {Niemiec}, {Oakes}, {O'Brien}, {Odaka}, {Ohm}, {Ostrowski},
  {Oya}, {Padovani}, {Panter}, {Parsons}, {Pekeur}, {Pelletier}, {Perennes},
  {Petrucci}, {Peyaud}, {Piel}, {Pita}, {Poireau}, {Poon}, {Prokhorov},
  {Prokoph}, {P{\"u}hlhofer}, {Punch}, {Quirrenbach}, {Raab}, {Rauth},
  {Reimer}, {Reimer}, {Renaud}, {de los Reyes}, {Rieger}, {Rinchiuso},
  {Romoli}, {Rowell}, {Rudak}, {Rulten}, {Sahakian}, {Saito}, {Sanchez},
  {Santangelo}, {Sasaki}, {Schlickeiser}, {Sch{\"u}ssler}, {Schulz},
  {Schwanke}, {Schwemmer}, {Seglar-Arroyo}, {Settimo}, {Seyffert}, {Shafi},
  {Shilon}, {Shiningayamwe}, {Simoni}, {Sol}, {Spanier}, {Spir-Jacob},
  {Stawarz}, {Steenkamp}, {Stegmann}, {Steppa}, {Sushch}, {Takahashi},
  {Tavernet}, {Tavernier}, {Taylor}, {Terrier}, {Tibaldo}, {Tiziani},
  {Tluczykont}, {Trichard}, {Tsirou}, {Tsuji}, {Tuffs}, {Uchiyama}, {van der
  Walt}, {van Eldik}, {van Rensburg}, {van Soelen}, {Vasileiadis}, {Veh},
  {Venter}, {Viana}, {Vincent}, {Vink}, {Voisin}, {V{\"o}lk}, {Vuillaume},
  {Wadiasingh}, {Wagner}, {Wagner}, {Wagner}, {White}, {Wierzcholska},
  {Willmann}, {W{\"o}rnlein}, {Wouters}, {Yang}, {Zaborov}, {Zacharias},
  {Zanin}, {Zdziarski}, {Zech}, {Zefi}, {Ziegler}, {Zorn}, {{\.Z}ywucka},
  {H.~E.~S.~S. Collaboration}, {Fender}, {Broderick}, {Rowlinson}, {Wijers},
  {Stewart}, {ter Veen}, {Shulevski}, {LOFAR Collaboration}, {Kavic},
  {Simonetti}, {League}, {Tsai}, {Obenberger}, {Nathaniel}, {Taylor}, {Dowell},
  {Liebling}, {Estes}, {Lippert}, {Sharma}, {Vincent}, {Farella}, {Wavelength
  Array}, {Abeysekara}, {Albert}, {Alfaro}, {Alvarez}, {Arceo},
  {Arteaga-Vel{\'a}zquez}, {Avila Rojas}, {Ayala Solares}, {Barber}, {Becerra
  Gonzalez}, {Becerril}, {Belmont-Moreno}, {BenZvi}, {Berley}, {Bernal},
  {Braun}, {Brisbois}, {Caballero-Mora}, {Capistr{\'a}n}, {Carrami{\~n}ana},
  {Casanova}, {Castillo}, {Cotti}, {Cotzomi}, {Couti{\~n}o de Le{\'o}n}, {De
  Le{\'o}n}, {De la Fuente}, {Diaz Hernandez}, {Dichiara}, {Dingus},
  {DuVernois}, {D{\'\i}az-V{\'e}lez}, {Ellsworth}, {Engel},
  {Enr{\'\i}quez-Rivera}, {Fiorino}, {Fleischhack}, {Fraija},
  {Garc{\'\i}a-Gonz{\'a}lez}, {Garfias}, {Gerhardt}, {Gonz{\~o}lez Mu{\~n}oz},
  {Gonz{\'a}lez}, {Goodman}, {Hampel-Arias}, {Harding}, {Hernandez},
  {Hernandez-Almada}, {Hona}, {H{\"u}ntemeyer}, {Iriarte}, {Jardin-Blicq},
  {Joshi}, {Kaufmann}, {Kieda}, {Lara}, {Lauer}, {Lennarz}, {Le{\'o}n Vargas},
  {Linnemann}, {Longinotti}, {Raya}, {Luna-Garc{\'\i}a}, {L{\'o}pez-Coto},
  {Malone}, {Marinelli}, {Martinez}, {Martinez-Castellanos},
  {Mart{\'\i}nez-Castro}, {Mart{\'\i}nez-Huerta}, {Matthews},
  {Miranda-Romagnoli}, {Moreno}, {Mostaf{\'a}}, {Nellen}, {Newbold}, {Nisa},
  {Noriega-Papaqui}, {Pelayo}, {Pretz}, {P{\'e}rez-P{\'e}rez}, {Ren}, {Rho},
  {Rivi{\`e}re}, {Rosa-Gonz{\'a}lez}, {Rosenberg}, {Ruiz-Velasco}, {Salazar},
  {Salesa Greus}, {Sandoval}, {Schneider}, {Schoorlemmer}, {Sinnis}, {Smith},
  {Springer}, {Surajbali}, {Tibolla}, {Tollefson}, {Torres}, {Ukwatta},
  {Weisgarber}, {Westerhoff}, {Wisher}, {Wood}, {Yapici}, {Yodh}, {Younk},
  {Zhou}, {{\'A}lvarez}, {HAWC Collaboration}, {Aab}, {Abreu}, {Aglietta},
  {Albuquerque}, {Albury}, {Allekotte}, {Almela}, {Alvarez Castillo},
  {Alvarez-Mu{\~n}iz}, {Anastasi}, {Anchordoqui}, {Andrada}, {Andringa},
  {Aramo}, {Arsene}, {Asorey}, {Assis}, {Avila}, {Badescu}, {Balaceanu},
  {Barbato}, {Barreira Luz}, {Becker}, {Bellido}, {Berat}, {Bertaina},
  {Bertou}, {Biermann}, {Biteau}, {Blaess}, {Blanco}, {Blazek}, {Bleve},
  {Boh{\'a}{\v{c}}ov{\'a}}, {Bonifazi}, {Borodai}, {Botti}, {Brack}, {Brancus},
  {Bretz}, {Bridgeman}, {Briechle}, {Buchholz}, {Bueno}, {Buitink}, {Buscemi},
  {Caballero-Mora}, {Caccianiga}, {Cancio}, {Canfora}, {Caruso}, {Castellina},
  {Catalani}, {Cataldi}, {Cazon}, {Chavez}, {Chinellato}, {Chudoba}, {Clay},
  {Cobos Cerutti}, {Colalillo}, {Coleman}, {Collica}, {Coluccia},
  {Concei{\c{c}}{\~a}o}, {Consolati}, {Contreras}, {Cooper}, {Coutu},
  {Covault}, {Cronin}, {D'Amico}, {Daniel}, {Dasso}, {Daumiller}, {Dawson},
  {Day}, {de Almeida}, {de Jong}, {De Mauro}, {de Mello Neto}, {De Mitri}, {de
  Oliveira}, {de Souza}, {Debatin}, {Deligny}, {D{\'\i}az Castro}, {Diogo},
  {Dobrigkeit}, {D'Olivo}, {Dorosti}, {Dos Anjos}, {Dova}, {Dundovic}, {Ebr},
  {Engel}, {Erdmann}, {Erfani}, {Escobar}, {Espadanal}, {Etchegoyen}, {Falcke},
  {Farmer}, {Farrar}, {Fauth}, {Fazzini}, {Feldbusch}, {Fenu}, {Fick},
  {Figueira}, {Filip{\v{c}}i{\v{c}}}, {Freire}, {Fujii}, {Fuster},
  {Ga{\"\i}or}, {Garc{\'\i}a}, {Gat{\'e}}, {Gemmeke}, {Gherghel-Lascu}, {Ghia},
  {Giaccari}, {Giammarchi}, {Giller}, {G{\l}as}, {Glaser}, {Golup}, {G{\'o}mez
  Berisso}, {G{\'o}mez Vitale}, {Gonz{\'a}lez}, {Gorgi}, {Gottowik}, {Grillo},
  {Grubb}, {Guarino}, {Guedes}, {Halliday}, {Hampel}, {Hansen}, {Harari},
  {Harrison}, {Harvey}, {Haungs}, {Hebbeker}, {Heck}, {Heimann}, {Herve},
  {Hill}, {Hojvat}, {Holt}, {Homola}, {H{\"o}randel}, {Horvath},
  {Hrabovsk{\'y}}, {Huege}, {Hulsman}, {Insolia}, {Isar}, {Jandt}, {Johnsen},
  {Josebachuili}, {Jurysek}, {K{\"a}{\"a}p{\"a}}, {Kampert}, {Keilhauer},
  {Kemmerich}, {Kemp}, {Kieckhafer}, {Klages}, {Kleifges}, {Kleinfeller},
  {Krause}, {Krohm}, {Kuempel}, {Kukec Mezek}, {Kunka}, {Kuotb Awad}, {Lago},
  {LaHurd}, {Lang}, {Lauscher}, {Legumina}, {Leigui de Oliveira},
  {Letessier-Selvon}, {Lhenry-Yvon}, {Link}, {Lo Presti}, {Lopes}, {L{\'o}pez},
  {L{\'o}pez Casado}, {Lorek}, {Luce}, {Lucero}, {Malacari}, {Mallamaci},
  {Mandat}, {Mantsch}, {Mariazzi}, {Maris}, {Marsella}, {Martello}, {Martinez},
  {Mart{\'\i}nez Bravo}, {Mas{\'\i}as Meza}, {Mathes}, {Mathys}, {Matthews},
  {Matthiae}, {Mayotte}, {Mazur}, {Medina}, {Medina-Tanco}, {Melo},
  {Menshikov}, {Merenda}, {Michal}, {Micheletti}, {Middendorf}, {Miramonti},
  {Mitrica}, {Mockler}, {Mollerach}, {Montanet}, {Morello}, {Morlino},
  {M{\"u}ller}, {M{\"u}ller}, {Muller}, {M{\"u}ller}, {Mussa}, {Naranjo},
  {Nguyen}, {Niculescu-Oglinzanu}, {Niechciol}, {Niemietz}, {Niggemann},
  {Nitz}, {Nosek}, {Novotny}, {No{\v{z}}ka}, {N{\'u}{\~n}ez}, {Oikonomou},
  {Olinto}, {Palatka}, {Pallotta}, {Papenbreer}, {Parente}, {Parra}, {Paul},
  {Pech}, {Pedreira}, {P{\c{e}}kala}, {Pe{\~n}a-Rodriguez}, {Pereira},
  {Perlin}, {Perrone}, {Peters}, {Petrera}, {Phuntsok}, {Pierog}, {Pimenta},
  {Pirronello}, {Platino}, {Plum}, {Poh}, {Porowski}, {Prado}, {Privitera},
  {Prouza}, {Quel}, {Querchfeld}, {Quinn}, {Ramos-Pollan}, {Rautenberg},
  {Ravignani}, {Ridky}, {Riehn}, {Risse}, {Ristori}, {Rizi}, {Rodrigues de
  Carvalho}, {Rodriguez Fernandez}, {Rodriguez Rojo}, {Roncoroni}, {Roth},
  {Roulet}, {Rovero}, {Ruehl}, {Saffi}, {Saftoiu}, {Salamida}, {Salazar},
  {Saleh}, {Salina}, {S{\'a}nchez}, {Sanchez-Lucas}, {Santos}, {Santos},
  {Sarazin}, {Sarmento}, {Sarmiento-Cano}, {Sato}, {Schauer}, {Scherini},
  {Schieler}, {Schimp}, {Schmidt}, {Scholten}, {Schov{\'a}nek}, {Schr{\"o}der},
  {Schr{\"o}der}, {Schulz}, {Schumacher}, {Sciutto}, {Segreto}, {Shadkam},
  {Shellard}, {Sigl}, {Silli}, {{\v{S}}m{\'\i}da}, {Snow}, {Sommers},
  {Sonntag}, {Soriano}, {Squartini}, {Stanca}, {Stani{\v{c}}}, {Stasielak},
  {Stassi}, {Stolpovskiy}, {Strafella}, {Streich}, {Suarez},
  {Suarez-Dur{\'a}n}, {Sudholz}, {Suomij{\"a}rvi}, {Supanitsky},
  {{\v{S}}up{\'\i}k}, {Swain}, {Szadkowski}, {Taboada}, {Taborda},
  {Timmermans}, {Todero Peixoto}, {Tomankova}, {Tom{\'e}}, {Torralba Elipe},
  {Travnicek}, {Trini}, {Tueros}, {Ulrich}, {Unger}, {Urban}, {Vald{\'e}s
  Galicia}, {Vali{\~n}o}, {Valore}, {van Aar}, {van Bodegom}, {van den Berg},
  {van Vliet}, {Varela}, {Vargas C{\'a}rdenas}, {V{\'a}zquez}, {Veberi{\v{c}}},
  {Ventura}, {Vergara Quispe}, {Verzi}, {Vicha}, {Villase{\~n}or}, {Vorobiov},
  {Wahlberg}, {Wainberg}, {Walz}, {Watson}, {Weber}, {Weindl}, {Wiede{\'n}ski},
  {Wiencke}, {Wilczy{\'n}ski}, {Wirtz}, {Wittkowski}, {Wundheiler}, {Yang},
  {Yushkov}, {Zas}, {Zavrtanik}, {Zavrtanik}, {Zepeda}, {Zimmermann},
  {Ziolkowski}, {Zong}, {Zuccarello}, {Pierre Auger Collaboration}, {Kim},
  {Schulze}, {Bauer}, {Corral-Santana}, {de Gregorio-Monsalvo},
  {Gonz{\'a}lez-L{\'o}pez}, {Hartmann}, {Ishwara-Chandra}, {Mart{\'\i}n},
  {Mehner}, {Misra}, {Micha{\l}owski}, {Resmi}, {ALMA Collaboration}, {Paragi},
  {Agudo}, {An}, {Beswick}, {Casadio}, {Frey}, {Jonker}, {Kettenis}, {Marcote},
  {Moldon}, {Szomoru}, {van Langevelde}, {Yang}, {Euro VLBI Team}, {Cwiek},
  {Cwiok}, {Czyrkowski}, {Dabrowski}, {Kasprowicz}, {Mankiewicz}, {Nawrocki},
  {Opiela}, {Piotrowski}, {Wrochna}, {Zaremba}, {{\.Z}arnecki}, {Pi of Sky
  Collaboration}, {Haggard}, {Nynka}, {Ruan}, {Chandra Team at McGill
  University}, {Bland}, {Booler}, {Devillepoix}, {de Gois}, {Hancock}, {Howie},
  {Paxman}, {Sansom}, {Towner}, {Desert Fireball Network}, {Tonry}, {Coughlin},
  {Stubbs}, {Denneau}, {Heinze}, {Stalder}, {Weiland}, {ATLAS}, {Eatough},
  {Kramer}, {Kraus}, {Time Resolution Universe Survey}, {Troja}, {Piro},
  {Becerra Gonz{\'a}lez}, {Butler}, {Fox}, {Khandrika}, {Kutyrev}, {Lee},
  {Ricci}, {Ryan}, {S{\'a}nchez-Ram{\'\i}rez}, {Veilleux}, {Watson},
  {Wieringa}, {Burgess}, {van Eerten}, {Fontes}, {Fryer}, {Korobkin},
  {Wollaeger}, {RIMAS}, {RATIR}, {Camilo}, {Foley}, {Goedhart}, {Makhathini},
  {Oozeer}, {Smirnov}, {Fender}, {Woudt}, \& {South
  Africa/MeerKAT}}]{Abbott2017ApJ848L12A}
{Abbott}, B.~P., {Abbott}, R., {Abbott}, T.~D., {et~al.} 2017, \apjl, 848, L12,
  \dodoi{10.3847/2041-8213/aa91c9}

\bibitem[{{Ackermann} {et~al.}(2011){Ackermann}, {Ajello}, {Asano}, {Axelsson},
  {Baldini}, {Ballet}, {Barbiellini}, {Baring}, {Bastieri}, {Bechtol},
  {Bellazzini}, {Berenji}, {Bhat}, {Bissaldi}, {Blandford}, {Bonamente},
  {Borgland}, {Bouvier}, {Bregeon}, {Brez}, {Briggs}, {Brigida}, {Bruel},
  {Buehler}, {Buson}, {Caliandro}, {Cameron}, {Caraveo}, {Carrigan},
  {Casandjian}, {Cecchi}, {{\c{C}}elik}, {Chaplin}, {Charles}, {Chekhtman},
  {Chiang}, {Ciprini}, {Claus}, {Cohen-Tanugi}, {Connaughton}, {Conrad},
  {Cutini}, {Dermer}, {de Angelis}, {de Palma}, {Dingus}, {Silva}, {Drell},
  {Dubois}, {Favuzzi}, {Fegan}, {Ferrara}, {Focke}, {Frailis}, {Fukazawa},
  {Funk}, {Fusco}, {Gargano}, {Gasparrini}, {Gehrels}, {Germani}, {Giglietto},
  {Giordano}, {Giroletti}, {Glanzman}, {Godfrey}, {Goldstein}, {Granot},
  {Greiner}, {Grenier}, {Grove}, {Guiriec}, {Hadasch}, {Hanabata}, {Harding},
  {Hayashi}, {Hayashida}, {Hays}, {Horan}, {Hughes}, {Itoh}, {J{\'o}hannesson},
  {Johnson}, {Johnson}, {Kamae}, {Katagiri}, {Kataoka}, {Kippen},
  {Kn{\"o}dlseder}, {Kocevski}, {Kouveliotou}, {Kuss}, {Lande}, {Latronico},
  {Lee}, {Llena Garde}, {Longo}, {Loparco}, {Lovellette}, {Lubrano}, {Makeev},
  {Mazziotta}, {McBreen}, {McEnery}, {McGlynn}, {Meegan}, {Mehault},
  {M{\'e}sz{\'a}ros}, {Michelson}, {Mizuno}, {Monte}, {Monzani}, {Moretti},
  {Morselli}, {Moskalenko}, {Murgia}, {Nakajima}, {Nakamori}, {Naumann-Godo},
  {Nishino}, {Nolan}, {Norris}, {Nuss}, {Ohno}, {Ohsugi}, {Okumura}, {Omodei},
  {Orlando}, {Ormes}, {Ozaki}, {Paciesas}, {Paneque}, {Panetta}, {Parent},
  {Pelassa}, {Pepe}, {Pesce-Rollins}, {Petrosian}, {Piron}, {Porter}, {Preece},
  {Racusin}, {Rain{\`o}}, {Rando}, {Rau}, {Razzano}, {Razzaque}, {Reimer},
  {Reimer}, {Reposeur}, {Reyes}, {Ripken}, {Ritz}, {Roth}, {Ryde},
  {Sadrozinski}, {Sander}, {Scargle}, {Schalk}, {Sgr{\`o}}, {Siskind}, {Smith},
  {Spandre}, {Spinelli}, {Stamatikos}, {Stecker}, {Strickman}, {Suson},
  {Tajima}, {Takahashi}, {Tanaka}, {Tanaka}, {Thayer}, {Thayer}, {Tibaldo},
  {Tierney}, {Toma}, {Torres}, {Tosti}, {Tramacere}, {Uchiyama}, {Uehara},
  {Usher}, {Vandenbroucke}, {van der Horst}, {Vasileiou}, {Vilchez}, {Vitale},
  {von Kienlin}, {Waite}, {Wang}, {Wilson-Hodge}, {Winer}, {Wood}, {Wu},
  {Yamazaki}, {Yang}, {Ylinen}, \& {Ziegler}}]{Ackermann2011ApJ}
{Ackermann}, M., {Ajello}, M., {Asano}, K., {et~al.} 2011, \apj, 729, 114,
  \dodoi{10.1088/0004-637X/729/2/114}

\bibitem[{{Arnaud}(1996)}]{xspec1996ASPC}
{Arnaud}, K.~A. 1996, in Astronomical Society of the Pacific Conference Series,
  Vol. 101, Astronomical Data Analysis Software and Systems V, ed. G.~H.
  {Jacoby} \& J.~{Barnes}, 17

\bibitem[{{Chen} {et~al.}(2018){Chen}, {Liu}, \& {Wang}}]{2018MNRAS.478..749C}
{Chen}, Y., {Liu}, R.-Y., \& {Wang}, X.-Y. 2018, \mnras, 478, 749,
  \dodoi{10.1093/mnras/sty1171}

\bibitem[{{di Lalla} {et~al.}(2022){di Lalla}, {Axelsson}, {Arimoto}, {Omodei},
  \& {Crnogoreeviae}}]{2022GCNlat}
{di Lalla}, N., {Axelsson}, M., {Arimoto}, M., {Omodei}, N., \&
  {Crnogoreeviae}, M. 2022, GRB Coordinates Network, 32283, 1

\bibitem[{{Fenimore} {et~al.}(1993){Fenimore}, {Epstein}, \&
  {Ho}}]{Fenimore1993A&AS}
{Fenimore}, E.~E., {Epstein}, R.~I., \& {Ho}, C. 1993, \aaps, 97, 59

\bibitem[{{Fermi GBM Team}(2022)}]{2022GCNgbm1}
{Fermi GBM Team}. 2022, GRB Coordinates Network, 32278, 1

\bibitem[{{Frederiks} {et~al.}(2022){Frederiks}, {Lysenko}, {Ridnaya},
  {Svinkin}, {Tsvetkova}, {Ulanov}, {Cline}, \& {Konus-Wind Team}}]{2022GCNkw}
{Frederiks}, D., {Lysenko}, A., {Ridnaya}, A., {et~al.} 2022, GRB Coordinates
  Network, 32295, 1

\bibitem[{{Galama} {et~al.}(1998){Galama}, {Vreeswijk}, {van Paradijs},
  {Kouveliotou}, {Augusteijn}, {B{\"o}hnhardt}, {Brewer}, {Doublier},
  {Gonzalez}, {Leibundgut}, {Lidman}, {Hainaut}, {Patat}, {Heise}, {in't Zand},
  {Hurley}, {Groot}, {Strom}, {Mazzali}, {Iwamoto}, {Nomoto}, {Umeda},
  {Nakamura}, {Young}, {Suzuki}, {Shigeyama}, {Koshut}, {Kippen}, {Robinson},
  {de Wildt}, {Wijers}, {Tanvir}, {Greiner}, {Pian}, {Palazzi}, {Frontera},
  {Masetti}, {Nicastro}, {Feroci}, {Costa}, {Piro}, {Peterson}, {Tinney},
  {Boyle}, {Cannon}, {Stathakis}, {Sadler}, {Begam}, \&
  {Ianna}}]{Galama1998Natur.395..670G}
{Galama}, T.~J., {Vreeswijk}, P.~M., {van Paradijs}, J., {et~al.} 1998, \nat,
  395, 670, \dodoi{10.1038/27150}

\bibitem[{{Gendre} {et~al.}(2013){Gendre}, {Stratta}, {Atteia}, {Basa},
  {Bo{\"e}r}, {Coward}, {Cutini}, {D'Elia}, {Howell}, {Klotz}, \&
  {Piro}}]{Gendre2013ApJ}
{Gendre}, B., {Stratta}, G., {Atteia}, J.~L., {et~al.} 2013, \apj, 766, 30,
  \dodoi{10.1088/0004-637X/766/1/30}

\bibitem[{{Gompertz} \& {Fruchter}(2017)}]{Gompertz2017ApJ}
{Gompertz}, B., \& {Fruchter}, A. 2017, \apj, 839, 49,
  \dodoi{10.3847/1538-4357/aa6629}

\bibitem[{{Granot} {et~al.}(2008){Granot}, {Cohen-Tanugi}, \&
  {Silva}}]{2008ApJ...677...92G}
{Granot}, J., {Cohen-Tanugi}, J., \& {Silva}, E. d. C.~e. 2008, \apj, 677, 92,
  \dodoi{10.1086/526414}

\bibitem[{{Hasco{\"e}t} {et~al.}(2012){Hasco{\"e}t}, {Daigne}, {Mochkovitch},
  \& {Vennin}}]{2012MNRAS.421..525H}
{Hasco{\"e}t}, R., {Daigne}, F., {Mochkovitch}, R., \& {Vennin}, V. 2012,
  \mnras, 421, 525, \dodoi{10.1111/j.1365-2966.2011.20332.x}

\bibitem[{{He} {et~al.}(2011){He}, {Wu}, {Toma}, {Wang}, \&
  {M{\'e}sz{\'a}ros}}]{2011ApJ...733...22H}
{He}, H.-N., {Wu}, X.-F., {Toma}, K., {Wang}, X.-Y., \& {M{\'e}sz{\'a}ros}, P.
  2011, \apj, 733, 22, \dodoi{10.1088/0004-637X/733/1/22}

\bibitem[{{Ioka} {et~al.}(2016){Ioka}, {Hotokezaka}, \& {Piran}}]{Ioka2016ApJ}
{Ioka}, K., {Hotokezaka}, K., \& {Piran}, T. 2016, \apj, 833, 110,
  \dodoi{10.3847/1538-4357/833/1/110}

\bibitem[{{Izzo} {et~al.}(2022){Izzo}, {D'Elia}, {de Ugarte Postigo}, {Fynbo},
  {Kann}, {Levan}, {Malesani}, {Saccardi}, {Thoene}, {Agui Fernandez}, {de
  Wet}, {Groot}, \& {Stargate Consortium}}]{2022GCN.32291....1I}
{Izzo}, L., {D'Elia}, V., {de Ugarte Postigo}, A., {et~al.} 2022, GRB
  Coordinates Network, 32291, 1

\bibitem[{{Kalantari} {et~al.}(2021){Kalantari}, {Ibrahim}, {Reza Rahimi
  Tabar}, \& {Rahvar}}]{Kalantari2021ApJ...922...77K}
{Kalantari}, Z., {Ibrahim}, A., {Reza Rahimi Tabar}, M., \& {Rahvar}, S. 2021,
  \apj, 922, 77, \dodoi{10.3847/1538-4357/ac1c06}

\bibitem[{{Kashiyama} {et~al.}(2013){Kashiyama}, {Nakauchi}, {Suwa}, {Yajima},
  \& {Nakamura}}]{Kashiyama2013ApJ}
{Kashiyama}, K., {Nakauchi}, D., {Suwa}, Y., {Yajima}, H., \& {Nakamura}, T.
  2013, \apj, 770, 8, \dodoi{10.1088/0004-637X/770/1/8}

\bibitem[{{Kouveliotou} {et~al.}(1993){Kouveliotou}, {Meegan}, {Fishman},
  {Bhat}, {Briggs}, {Koshut}, {Paciesas}, \&
  {Pendleton}}]{Kouveliotou1993ApJ...413L.101K}
{Kouveliotou}, C., {Meegan}, C.~A., {Fishman}, G.~J., {et~al.} 1993, \apjl,
  413, L101, \dodoi{10.1086/186969}

\bibitem[{{Krolik} \& {Pier}(1991)}]{Krolik1991ApJ}
{Krolik}, J.~H., \& {Pier}, E.~A. 1991, \apj, 373, 277, \dodoi{10.1086/170048}

\bibitem[{{Krolik} \& {Piran}(2011)}]{Krolik2011ApJ}
{Krolik}, J.~H., \& {Piran}, T. 2011, \apj, 743, 134,
  \dodoi{10.1088/0004-637X/743/2/134}

\bibitem[{{Kumar} \& {Barniol Duran}(2009)}]{2009MNRAS.400L..75K}
{Kumar}, P., \& {Barniol Duran}, R. 2009, \mnras, 400, L75,
  \dodoi{10.1111/j.1745-3933.2009.00766.x}

\bibitem[{{Levan} {et~al.}(2014){Levan}, {Tanvir}, {Starling}, {Wiersema},
  {Page}, {Perley}, {Schulze}, {Wynn}, {Chornock}, {Hjorth}, {Cenko},
  {Fruchter}, {O'Brien}, {Brown}, {Tunnicliffe}, {Malesani}, {Jakobsson},
  {Watson}, {Berger}, {Bersier}, {Cobb}, {Covino}, {Cucchiara}, {de Ugarte
  Postigo}, {Fox}, {Gal-Yam}, {Goldoni}, {Gorosabel}, {Kaper}, {Kr{\"u}hler},
  {Karjalainen}, {Osborne}, {Pian}, {S{\'a}nchez-Ram{\'\i}rez}, {Schmidt},
  {Skillen}, {Tagliaferri}, {Th{\"o}ne}, {Vaduvescu}, {Wijers}, \&
  {Zauderer}}]{Levan2014ApJ}
{Levan}, A.~J., {Tanvir}, N.~R., {Starling}, R.~L.~C., {et~al.} 2014, \apj,
  781, 13, \dodoi{10.1088/0004-637X/781/1/13}

\bibitem[{{Lin} {et~al.}(2022){Lin}, {Li}, {Gao}, {Lin}, {Zhang}, {Liu}, {Zou},
  {Zhang}, {Zhou}, {Li}, \& {Lan}}]{lin2022ApJ...931....4L}
{Lin}, S.-J., {Li}, A., {Gao}, H., {et~al.} 2022, \apj, 931, 4,
  \dodoi{10.3847/1538-4357/ac6505}

\bibitem[{{Lithwick} \& {Sari}(2001)}]{2001ApJ...555..540L}
{Lithwick}, Y., \& {Sari}, R. 2001, \apj, 555, 540, \dodoi{10.1086/321455}

\bibitem[{{Mao}(1992)}]{mao1992ApJ}
{Mao}, S. 1992, \apjl, 389, L41, \dodoi{10.1086/186344}

\bibitem[{{Mazets} {et~al.}(1981){Mazets}, {Golenetskii}, {Ilinskii}, {Panov},
  {Aptekar}, {Gurian}, {Proskura}, {Sokolov}, {Sokolova}, \&
  {Kharitonova}}]{1981Ap&SS803M}
{Mazets}, E.~P., {Golenetskii}, S.~V., {Ilinskii}, V.~N., {et~al.} 1981, \apss,
  80, 3, \dodoi{10.1007/BF00649140}

\bibitem[{{Meegan} {et~al.}(2009){Meegan}, {Lichti}, {Bhat}, {Bissaldi},
  {Briggs}, {Connaughton}, {Diehl}, {Fishman}, {Greiner}, {Hoover}, {van der
  Horst}, {von Kienlin}, {Kippen}, {Kouveliotou}, {McBreen}, {Paciesas},
  {Preece}, {Steinle}, {Wallace}, {Wilson}, \& {Wilson-Hodge}}]{Meegan2009ApJ}
{Meegan}, C., {Lichti}, G., {Bhat}, P.~N., {et~al.} 2009, \apj, 702, 791,
  \dodoi{10.1088/0004-637X/702/1/791}

\bibitem[{{Nakauchi} {et~al.}(2013){Nakauchi}, {Kashiyama}, {Suwa}, \&
  {Nakamura}}]{Nakauchi2013ApJ}
{Nakauchi}, D., {Kashiyama}, K., {Suwa}, Y., \& {Nakamura}, T. 2013, \apj, 778,
  67, \dodoi{10.1088/0004-637X/778/1/67}

\bibitem[{{Norris} {et~al.}(1984){Norris}, {Cline}, {Desai}, \&
  {Teegarden}}]{Norris1984Natur.308..434N}
{Norris}, J.~P., {Cline}, T.~L., {Desai}, U.~D., \& {Teegarden}, B.~J. 1984,
  \nat, 308, 434, \dodoi{10.1038/308434a0}

\bibitem[{{Nunes} {et~al.}(2017){Nunes}, {Pan}, {Saridakis}, \&
  {Abreu}}]{2017JCAP...01..005N}
{Nunes}, R.~C., {Pan}, S., {Saridakis}, E.~N., \& {Abreu}, E. M.~C. 2017,
  \jcap, 2017, 005, \dodoi{10.1088/1475-7516/2017/01/005}

\bibitem[{{Paynter} {et~al.}(2021){Paynter}, {Webster}, \&
  {Thrane}}]{Paynter2021NatAs...5..560P}
{Paynter}, J., {Webster}, R., \& {Thrane}, E. 2021, Nature Astronomy, 5, 560,
  \dodoi{10.1038/s41550-021-01307-1}

\bibitem[{{Peng} {et~al.}(2016){Peng}, {Tang}, \& {Wang}}]{2016ApJ...825...47P}
{Peng}, F.-K., {Tang}, Q.-W., \& {Wang}, X.-Y. 2016, \apj, 825, 47,
  \dodoi{10.3847/0004-637X/825/1/47}

\bibitem[{{Perna} {et~al.}(2018){Perna}, {Lazzati}, \&
  {Cantiello}}]{Perna2018ApJ}
{Perna}, R., {Lazzati}, D., \& {Cantiello}, M. 2018, \apj, 859, 48,
  \dodoi{10.3847/1538-4357/aabcc1}

\bibitem[{{Raman} {et~al.}(2022){Raman}, {Tohuvavohu}, {DeLaunay}, \&
  {Kennea}}]{2022GCNswift}
{Raman}, G., {Tohuvavohu}, A., {DeLaunay}, J., \& {Kennea}, J.~A. 2022, GRB
  Coordinates Network, 32287, 1

\bibitem[{{Roberts} {et~al.}(2022){Roberts}, {Hristov}, {Meegan}, \& {Fermi
  Gamma-ray Burst Monitor Team}}]{2022GCNgbm2}
{Roberts}, O.~J., {Hristov}, B., {Meegan}, C., \& {Fermi Gamma-ray Burst
  Monitor Team}. 2022, GRB Coordinates Network, 32288, 1

\bibitem[{{Scargle} {et~al.}(2013){Scargle}, {Norris}, {Jackson}, \&
  {Chiang}}]{2013ApJ...764..167S}
{Scargle}, J.~D., {Norris}, J.~P., {Jackson}, B., \& {Chiang}, J. 2013, \apj,
  764, 167, \dodoi{10.1088/0004-637X/764/2/167}

\bibitem[{{Schwarz}(1978)}]{1978AnSta...6..461S}
{Schwarz}, G. 1978, Annals of Statistics, 6, 461

\bibitem[{{Stratta} {et~al.}(2013){Stratta}, {Gendre}, {Atteia}, {Bo{\"e}r},
  {Coward}, {De Pasquale}, {Howell}, {Klotz}, {Oates}, \&
  {Piro}}]{Stratta2013ApJ}
{Stratta}, G., {Gendre}, B., {Atteia}, J.~L., {et~al.} 2013, \apj, 779, 66,
  \dodoi{10.1088/0004-637X/779/1/66}

\bibitem[{{Tang} {et~al.}(2015){Tang}, {Peng}, {Wang}, \&
  {Tam}}]{2015ApJ...806..194T}
{Tang}, Q.-W., {Peng}, F.-K., {Wang}, X.-Y., \& {Tam}, P.-H.~T. 2015, \apj,
  806, 194, \dodoi{10.1088/0004-637X/806/2/194}

\bibitem[{{Treu}(2010)}]{2010ARA&A..48...87T}
{Treu}, T. 2010, \araa, 48, 87, \dodoi{10.1146/annurev-astro-081309-130924}

\bibitem[{{Veres} {et~al.}(2021){Veres}, {Bhat}, {Fraija}, \&
  {Lesage}}]{Veres2021ApJ...921L..30V}
{Veres}, P., {Bhat}, N., {Fraija}, N., \& {Lesage}, S. 2021, \apjl, 921, L30,
  \dodoi{10.3847/2041-8213/ac2ee6}

\bibitem[{{Vianello} {et~al.}(2018){Vianello}, {Gill}, {Granot}, {Omodei},
  {Cohen-Tanugi}, \& {Longo}}]{Vianello2018ApJ}
{Vianello}, G., {Gill}, R., {Granot}, J., {et~al.} 2018, \apj, 864, 163,
  \dodoi{10.3847/1538-4357/aad6ea}

\bibitem[{{Wang} {et~al.}(2010){Wang}, {He}, {Li}, {Wu}, \&
  {Dai}}]{2010ApJ...712.1232W}
{Wang}, X.-Y., {He}, H.-N., {Li}, Z., {Wu}, X.-F., \& {Dai}, Z.-G. 2010, \apj,
  712, 1232, \dodoi{10.1088/0004-637X/712/2/1232}

\bibitem[{{Wang} {et~al.}(2021){Wang}, {Jiang}, {Li}, {Ren}, {Tang}, {Zhou},
  {Liang}, \& {Fan}}]{Wang2021ApJ...918L..34W}
{Wang}, Y., {Jiang}, L.-Y., {Li}, C.-K., {et~al.} 2021, \apjl, 918, L34,
  \dodoi{10.3847/2041-8213/ac1ff9}

\bibitem[{{Yang} {et~al.}(2021){Yang}, {L{\"u}}, {Yuan}, {Rice}, {Zhang},
  {Zhang}, \& {Liang}}]{Yang2021ApJ...921L..29Y}
{Yang}, X., {L{\"u}}, H.-J., {Yuan}, H.-Y., {et~al.} 2021, \apjl, 921, L29,
  \dodoi{10.3847/2041-8213/ac2f39}

\bibitem[{{Zhang}(2018)}]{2018pgrb.book.....Z}
{Zhang}, B. 2018, {The Physics of Gamma-Ray Bursts},
  \dodoi{10.1017/9781139226530}

\bibitem[{{Zou} {et~al.}(2019){Zou}, {Zhou}, {Xie}, {Zhang}, {L{\"u}}, {Zhong},
  {Wang}, \& {Liang}}]{Zou2019ApJ}
{Zou}, L., {Zhou}, Z.-M., {Xie}, L., {et~al.} 2019, \apj, 877, 153,
  \dodoi{10.3847/1538-4357/ab17dc}

\end{thebibliography}
\bibliographystyle{aasjournal}

\newpage
\begin{table}[ht!]
\caption{Spectral fitting results of the GBM/LAT emission for the two episodes.}
\begin{center}
    \begin{tabular}{lcccc}
        \hline\hline 
       Model &  Band & CPL & Band+PL & CPL+PL\\ \hline\hline\\
        &   & \textbf{Episode \uppercase\expandafter{\romannumeral1}: $T_0$-3.08--$T_0$+288.26 s} &  & \\\\
\hline       
Band fuction & & & & \\
$\alpha$ & $-0.89\pm0.04$ &  & $-0.88\pm0.07$ & \\
$\beta$   & $-2.45\pm0.02$ &  & $-2.45\pm0.04$ & \\
$E_{p}$ (keV)  & $334.08\pm33.32$ &  & $327.936\pm58.84$ & \\
\hline
$CPL$& & & & \\
$\lambda$  &  & $0.91\pm0.03$ &  & $0.73\pm0.10$\\
$E_c$ (keV)  &  & $361.43\pm31.58$ &  & $286.96\pm38.00$\\
\hline
Powerlaw & & & & \\
Index & &  & $1.92\pm0.05$ &$1.92\pm0.05$\\
\hline
C-stat/dof  & 766.06/322 & 2243.42/323 & 752.30/320 & 752.39/321\\
BIC  & 789.21 & 2260.78 & 787.02 & 781.32\\
\hline\hline\\
  &  &\textbf{Episode \uppercase\expandafter{\romannumeral2}: $T_0$+907.15--$T_0$+1072.53 s}  &  & \\\\
\hline
Band fuction & & & & \\
$\alpha$  & $1.05\pm0.05$ &  &  & \\
$\beta$   & $<-9.27$ &  &  & \\
$E_{p}$ (keV)  & $247.77\pm66.45$ &  &  & \\
\hline
$CPL$& & & & \\
$\lambda$  & & $1.06\pm0.11$  &  \\
$E_c$ (keV)  & & $248.64\pm66.58$ &  &   \\
\hline
Powerlaw & & & & \\
Index  & &  &  & \\
\hline\hline
C-stat/dof  & 508.83/340 & 508.83/341 &  & \\
BIC  & 532.19 & 526.35 &  & \\
\hline\hline
    \end{tabular}
    \end{center}
\label{tab:1}
\end{table}{}


\begin{table}[ht!]
\caption{Time-resolved spectral fitting results.}
\begin{center}
    \begin{tabular}{lccccc}
        \hline\hline 
       Time Interval (s)  & a (-3.08--129.53) & b (129.53--195.07) & c (195.07--288.26 )& d (907.15--996.75) & e (996.75--1072.53)\\
Preferred model  & Band & CPL+PL & Band & CPL & CPL\\ \hline\hline
$Band~fuction$ & { } & { }& { }\\
$\alpha$  & $-0.81\pm-0.06$ &  & $-1.79\pm0.12$\\
$\beta$  & $-2.69\pm 0.07$ &  & $-2.26\pm0.46$\\
$E_{p}$ (keV)  & $282.27\pm35.63$ &   & $9748.51\pm6635.92$\\
\hline\hline
$CPL$& & & \\
$\lambda$&  & $0.64\pm0.10$ &  & $1.13\pm0.13$ & $0.88\pm0.22$\\
$E_c$ (keV) &  & $267.35\pm34.75$ &  & $342.23\pm130.68$ & $146.54\pm53.07$\\
\hline
$Powerlaw$ &  & & \\
Index &   & $1.84\pm0.06$ & \\
\hline\hline
C-stat/dof  & 441.51/316 & 369.18/317 & 345.79/316 & 345.42/311 & 308.09/311 \\
BIC & 464.584 & 398.04 & 368.85 & 362.668 & 325.338\\
\hline
$\Delta_{BIC}$ &  & & \\
Band     &  0   & 4.67 & 0      & 39.71 & 5.76\\
CPL+PL  & 5.11 & 0     & 5.45  &       &  \\
\hline\hline
    \end{tabular}
    \end{center}
\label{tab:intervals}
\end{table}{}

\end{document}